\documentclass[notitlepage,pre,amsfonts,floatfix,superscriptaddress,twocolumn,longbibliography]{revtex4-2}
\usepackage[utf8]{inputenc}
\usepackage{mathbbol}
\usepackage{graphicx,color}
\usepackage{mciteplus} 
\usepackage{amsmath,amssymb,bm,amsthm}
\usepackage{mathtools}
\usepackage{simplewick}
\usepackage{bbding}
\usepackage{braket}
\usepackage{subfigure} 
\usepackage{mathrsfs}
\usepackage{graphicx}
\usepackage[graphicx]{realboxes}
\usepackage{color}
\usepackage{multirow}
\usepackage{appendix}
\usepackage{float}
\usepackage{braket}
\usepackage[hidelinks]{hyperref}
\usepackage[table]{xcolor}

\begin{document}
\title{Late-time universal distribution functions of observables in one-dimensional many-body quantum systems}

\author{I. Vallejo-Fabila}
\author{E. Jonathan Torres-Herrera}

\address{Instituto de Física, Benemérita Universidad Autónoma de Puebla, Puebla, 72570, México}



\begin{abstract}
We study the probability distribution function of the long-time values of observables being time-evolved by Hamiltonians modeling clean and disordered one-dimensional chains of many spin-1/2 particles. In particular, we analyze the return probability and its version for a completely extended initial state, the so-called spectral form factor. We complement our analysis with the spin autocorrelation and connected spin-spin correlation functions, both of interest in experiments with quantum simulators.  We show that the distribution function has a universal shape provided the central limit theorem holds. Explicitly, the shape is exponential for the return probability and spectral form factor, meanwhile it is Gaussian for the few-body observables.  We also discuss implications over the so-called many-body localization. Remarkably, our approach requires only a single sample of the dynamics and small system sizes, which could be quite advantageous when dealing specially with disordered systems. 
\end{abstract}


\maketitle

\section{Introduction}\label{Sec. I}

Probability distribution functions (pdfs) are of fundamental interest in different branches of science. In physics, in particular, pdfs are important not only for the purpose of computing mean values or expectation values of relevant physical quantities but also by themselves because they could provide fine information about fluctuations around the mean values of properties of a given system. For a comprehensive review on fluctuations in physical systems see, for instance, Ref.~\cite{Clusel2008}.

In the realm of quantum mechanics, the characterization of fluctuations in time domain has been useful to understand reversibility in generic chaotic and integrable systems~\cite{Peres1984}, to establish a Gaussian scenario for equilibration in noninteracting models~\cite{Venuti2010,Venuti2013}, where fluctuations can decay as the square root of the system size. Although later is was shown in Ref.~\cite{Zangara2013} that for many-body systems the fluctuations of generic observables decay exponentially with system size in both, integrable and chaotic systems. 

Late-time pdfs are also important in the context of thermalization and many-body localization (MBL).  Just some examples are: Late-time exponential probability distribution functions experimentally observed for the squared-contrast of a one-dimensional Bose gas and related to prethermalization and thermal equilibrium in Ref.~\cite{Gring2012}. Gaussian momentum distribution functions for noninteracting spinless fermions and hard-core bosons in one-dimensional (1D) lattices were observed in Refs.~\cite{Gramsch2012,HeSantos2013}. Gaussian distributions are experimentally obtained for the equilibrium value of a temporal autocorrelation function, number entropy and Hamming distance in Ref.~\cite{gong2021}. In numerical experiments employing the time dependence variational principle for matrix product states and machine learning, late-time pdfs where analyzed for the spin imbalance, entanglement entropy and so-called Schmidt gap~\cite{Doggen2018}. Pdfs were useful for the determination of the universality class of MBL under speckle disorder~\cite{maksymov2020}. Long-tails and bimodal probability distribution functions of the entanglement entropy near and at the MBL transition, respectively, were previously analyzed in Refs.~\cite{luitz2016long,Yu2016}.  

From our point of view a clear determination of the conditions under which the formerly found, experimentally and theoretically, pdfs of observables in quantum systems is missing. Our aim in this work is twofold, one is to generalize the Gaussian scenario established in Refs.~\cite{Venuti2010,Venuti2013} and the other one is to set, in a systematic way, the basic conditions that lead to universal late-time probability distribution functions of physical quantities and observables in 1D interacting quantum systems. We will show that the central limit theorem (CLT) is behind the universal late-time pdfs, and will also analyze the conditions under which this almost ubiquitous and undoubtedly celebrated theorem holds in the time evolutions generated by paradigmatic models of spin-1/2 particles, with or without interactions, clean or disordered, chaotic or integrable. In particular, we conjecture that a sufficient condition to have universal probability distribution functions is that of linearly uncorrelated identically distributed (u.i.d.) variables. Projections of the initial state in the energy eigenstates, as well as eigenvalues are the main characters playing a relevant role in the fulfillment of the CLT. Of course, when dealing with observables, matrix elements in energy representation could be also relevant for the fulfillment of the CLT.

Our analysis will employ different probes of the dynamics, namely, return probability, spectral form factor, spin autocorrelation function and connected spin-spin correlation function. We will put main emphasis in the first one, the return probability, this because it has been prominently studied since the early years of quantum mechanics. Let us give another nonexhaustive but larger list of examples: In the context of the time-energy uncertainty relation~\cite{Mandelstam1945,Bhattacharyya1983}, nonexponential late-time decays of a quasistationary state~\cite{Khalfin1958}, quantum Zeno's paradox~\cite{Misra1977}, quantum speed limits~\cite{Pfeifer1993,Ufink1993,Margolus1998,Giovannetti2003}, quantum energy flow~\cite{Schofield1995}, nonperiodic substitution potentials~\cite{de1999quantum}, connection with the time operator through a generalized Weyl relation~\cite{Miyamoto2001}, cosmology ~\cite{krauss2008late}, quantum walks and complex networks~\cite{xu2008,mulken2011,ampadu2012return,riascos2015}, dynamics of a thermofield double state~\cite{Campo2017,CampoSantos2018}, and matter-radiation interaction models like Dicke ~\cite{lerma2019} and Bose-Hubbard~\cite{delacruz2020}. The return probability has been also measured in experiments to observe time-resolved level repulsion of chaotic systems~\cite{Leviandier1986}, fluorescence experiments with molecules~\cite{Gruebele2004,rothe2006violation}, ultracold atoms in magneto-optic traps~\cite{wilkinson1997}, atom chip~\cite{gherardini2017}, and prethermalization in Floquet systems~\cite{singh2019}.   

The rest of the manuscript is organized as follows. Models and quantities are introduced in Secs.~\ref{sec:mod} and~\ref{sec:QO}, respectively. In Sec.~\ref{sec:dynamics} we provide an overview of the dynamics from the time it is initiated up to equilibrium.  In Sec.~\ref{sec:CLT} we present a justification of the universal distributions based on theoretical grounds. Results for the addressed systems are shown and discussed in Sec.~\ref{Sec:Results}. Conclusions are finally presented in Sec.~\ref{Sec:Conclusions}.   

\section{Models}\label{sec:mod}
We consider a system of spin-$1/2$ particles that can interact with nearest-neighbors in a 1D lattice. Particles can also be subjected to an on-site random potential. The general Hamiltonian describing such a system is
\begin{equation}\label{eq:Ham}
\mathcal{H} = \sum _{k} (S_{k}^{x}S_{k+1}^{x}+S_{k}^{y}S_{k+1}^{y} + \Delta S_{k}^{z}S_{k+1}^{z}) + \sum _{k=1}^{L}h_{k}S_{k}^{z}.
\end{equation}
With spin-$1/2$ operators $S_{k}^{x,y,z}$ acting on the spin located at the $k$-th site. We set $\hbar=1$. The anisotropy parameter $\Delta$ and the on-site potential amplitudes $h_k$ are tuned to obtain different instances of model~\eqref{eq:Ham}. Setting $\Delta=0$ together with $h_k=0$ leads to a noninteracting model, known as $XX$ model which is solved by applying a Jordan-Wigner transformation~\cite{Jordan_1928,Lieb1961}. With $\Delta=1$ and $h_k=0$ we obtain the isotropic $XXZ$ model with Ising-like interactions in $z$ direction, this model in also solvable but via the celebrated Bethe ansatz~\cite{Bethe1931,hulthen1938,des1962,yang1966one,kirillov1987I}. We arrive to a paradigmatic model for the MBL transition by setting again $\Delta=1$ but with $h_k$ as uncorrelated random numbers uniformly distributed in the interval $(-h,h)$, with $h$ as the disorder strength. A critical point $h_c$ has been identified, $3.75 \lesssim h_c $, but still without consensus about its precise location in the $h$-axis, see for instance Refs.~\cite{Oganesyan2007,Pal2010,Berkelbach2010,Kjall2014,Luitz2015,Devakul2015,Doggen2018,Sierant_2020}. For the $XX$ and $XXZ$ models we consider open boundaries conditions, meanwhile for the disordered model we use periodic boundary conditions, $\hat{S}_{L+1}=\hat{S}_{1}$, implying that in the first summation of Hamiltonian~\eqref{eq:Ham} we have the index $k$ running from $1$ to $L-1$ for the two former models and from $1$ to $L$ for the latter. 

Hamiltonian~\eqref{eq:Ham} conserves the total spin in the $z$-direction, ${\cal{S}}^z=\sum_k S_k^z$, it only moves an excitation (spin up) through the chain due to the $XX$ term which is also known as the flip-flop term. This allows to focus our study on the largest subspace with ${\cal{S}}^z=0$ and dimension ${\cal{N}}=L!/(L/2)!^{2}$. 

\section{Quantities and observables}\label{sec:QO}
In this section we describe the time-dependent quantities and observables employed in our study. We also present the level spacing distribution traditionally used as a probe of repulsion between adjacent energy levels, a fingerprint of quantum chaos.
\subsection{Return probability and spectral form factor}\label{ssub:rpsff}

The return probability, RP, also known as survival or nondecay probability is a dynamical quantity defined as the probability to find an initial state $\ket{\Psi(0)}$ at a future time $t$. It is given by
\begin{equation}\label{eq:rp}
\text{RP}(t) =  |S(t)|^{2}= \left| \displaystyle{\sum _{\alpha =1}^{\cal{N}}|c_{\alpha}^0|^{2}e^{-iE_{\alpha}t}} \right|^{2},
\end{equation}
where $S(t)=\bra{\Psi(0)}{\cal{U}}(t)\ket{\Psi(0)}$ is the return amplitude and ${\cal{U}}(t)=\exp(-i{\mathcal{H}}t)$ the unitary time evolution operator. For the Hamiltonian generating the dynamics we have the eigenvalue equation $\mathcal{H}\ket{\psi_\alpha} =E_{\alpha}\ket{\psi_\alpha}$, $\alpha=1,\,2,\,\dots,\,{\cal{N}}$. Meanwhile $c_{\alpha}^{0}=\braket{\psi_\alpha|\Psi(0)}$ are the projections of the initial state into the energy eigenstates.

When the initial state $\ket{\Psi (0)}$ is completely extended in the energy eigenbasis, its components goes as 
$1/\sqrt{\cal{N}}$ and from Eq.~\eqref{eq:rp} we recover the so-called spectral form factor,
\begin{equation}\label{eq:sff}
K(t)=\frac{1}{{\cal{N}}^{2}}\left|\sum_{\alpha=1}^{\cal{N}} e^{-iE_{\alpha}t} \right|^{2},
\end{equation}
where only energy eigenvalues are involved. 

\subsection{Spin autocorrelation function}
How close the spin configuration at time $t$ is to the one at $t=0$ can be measured through so-called spin autocorrelation function, given by
\begin{equation}\label{eqn:SAF}
I(t) = \frac{4}{L}\sum _{i=1}^{L}\braket{\Psi (0)|\hat{S}_{i}^{z}e^{i\hat{\mathcal{H}}t}\hat{S}_{i}^{z}e^{-i\hat{\mathcal{H}}t}|\Psi (0)}.
\end{equation}
For particular initial states, like a quantum Néel-like state, $I(t)$ is equivalent to the density imbalance measured in experimental platforms studying ultracold atoms~\cite{Schreiber2015,Bordia2017a}. 

\subsection{Connected spin-spin correlation function}
Another quantity also of interest in quantum simulators is the connected spin-spin correlation function, defined through 
\begin{equation}\label{eqn:CSSCF}
\begin{array}{ll}
C(t) & = \displaystyle{ \frac{4}{L}\sum _{k=1}^{L-1}\left[ \bra{\Psi (t)}S_{k}^{z}S_{k+1}^{z}\ket{\Psi (t)} \right.}\\
 &  \displaystyle{\left. \hspace{1.cm}-\bra{\Psi (t)}S_{k}^{z}\ket{\Psi (t)}\bra{\Psi (t)}S_{k+1}^{z}\ket{\Psi (t)} \right]}.
\end{array}
\end{equation}
It quantifies time-dependent correlations between neighboring spins in the chain. It has been measured, for instance, in experiments with trapped ions studying systems with long-range interactions~\cite{Richerme2014}.

\subsection{Level spacings}\label{sec:ls}
We present in this subsection the theoretical predictions for the distribution $P(s)$ of spacings between adjacent energy levels, $s_\alpha=(E_{\alpha+1}-E_{\alpha})/\delta E$, with $\delta E$ being the mean level spacing of the system. For integrable systems the level spacings typically follow an exponential distribution, they obey Poisson-like statistics~\cite{Berry1977},
\begin{equation}\label{eq:P}
 P_\text{P}(s)=\exp(-s). 
\end{equation}
No level repulsion is apparent from Eq.~\eqref{eq:P} since $P(s\to 0)\to 1$. Meanwhile spacings between energy levels of time-reversal invariant systems, just like the ones we deal with, that are chaotic follow statistics from the Gaussian orthogonal ensemble (GOE) of random matrix theory (RMT)~\cite{Bohigas1984},
\begin{equation}\label{eq:GOE}
 P_\text{GOE}(s)=\frac{\pi}{2}s\exp\left(-\dfrac{\pi}{4}s^2\right), 
\end{equation}
and level repulsion is manifested as indicated by the limit case $P(s\to 0)\to s$. We stress that $P(s)$ will be used just to signal the regime of the Hamiltonian, being integrable or chaotic (ergodic, thermalizing), and also to expose possible degeneracies of the energy spectrum, like in the $XX$ model.

\section{Dynamics overview}\label{sec:dynamics}
\begin{figure}[h]
\centering
\includegraphics[scale=0.45]{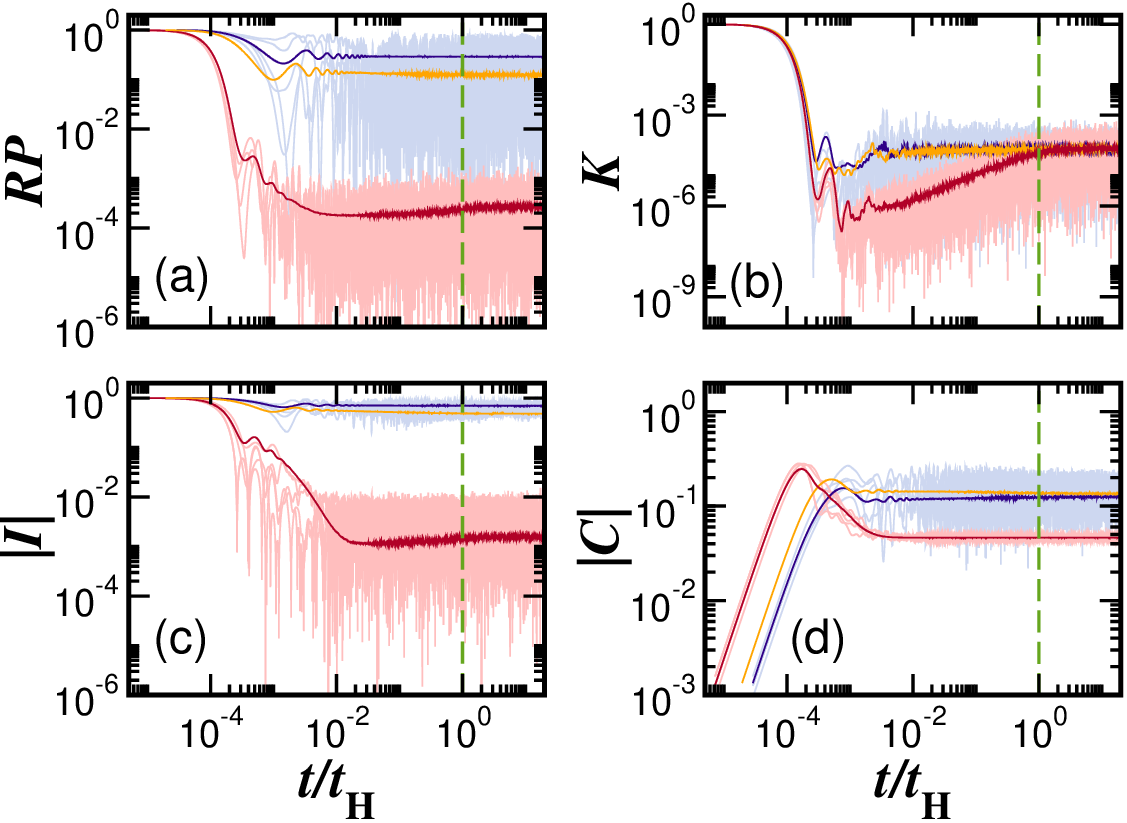}
\caption{Full-time dynamics of the quantities and observables treated in our study. Time evolution is generated by the disordered model in the ergodic phase with $h=0.5$ (red), $h=3.75$ (orange), and $h=6.0$ (blue). Return probability (a), spectral form factor (b), spin autocorrelation function (c), and connected spin-spin correlation function (d). Light-colored curves are results from five samples of the dynamics, meanwhile dark-colored curves correspond to an average over $\sim 10^4$ samples originated from different initial states and random realizations. System size is $L=16$.}
\label{fig:dyn}
\end{figure}

We focus on the late-time behavior of the dynamical quantities and observables introduced lines above, that is, beyond the so-called Heisenberg time, $t_\text{H}=2\pi/\delta E$. However, it is instructive to illustrate their time evolution from $t=0$ up to saturation. Figure~\ref{fig:dyn} depicts the full-time evolution generated by the disordered model with fixed number of sites $L=16$ for $\text{RP}(t)$ [Fig.~\ref{fig:dyn}(a)], $K(t)$ [Fig.~\ref{fig:dyn}(b)], $I(t)$ [Fig.~\ref{fig:dyn}(c)], and $C(t)$ [Fig.~\ref{fig:dyn}(d)]. Three different disorder strengths are considered, $h=0.5$ (red), $h=3.75$ (yellow), and $h=6.0$ (blue). Dark-colored curves are averages over samples obtained from $130$ disorder realizations and $78$ initial states with energy closest to zero, $E_0=\bra{\Psi(0)}{\mathcal{H}}\ket{\Psi(0)}\approx 0$. Then a total of $\sim 10^3$ samples were considered for the average.  Light-colored curves are results from five individual samples, i.e., each one representing the dynamics for a single initial state with energy closest to zero and only one disorder realization. Note that individual samples for $h=3.75$ are omitted, this is done only to avoid overcrowding of figure. For $K(t)$ only disorder realizations were considered. The individual samples present huge sample-to-sample fluctuations~\cite{Prange1997} that could hide or obscure some relevant aspects of the dynamics, as the ones shown by the averaged quantities~\cite{Schiulaz2020,TorresHerrera2020a,TorresHerrera2020b}. 

In fact, from the averages shown in Fig.~\ref{fig:dyn} a clear power-law decay is seen for the return probability and apparently also for the imbalance~\cite{Torres2015,luitz2016}, although for this last quantity dissipative dynamics includes a stretched exponential decay~\cite{Fischer2016,Levi2016}. The power-law decay is followed by the correlation hole~\cite{TorresH2017,Torres2018,Torres2019} which is visible for both quantities but also for the spectral form factor.  For $C(t)$ the correlation hole is not apparent but it is present, a deep close up is needed to observe it~\cite{lezama2021}. We note that also in Ref.~\cite{lezama2021} it was shown how these features depend on both, system size and kind of observable.  

\begin{figure}[h]
\centering
\includegraphics[scale=0.45]{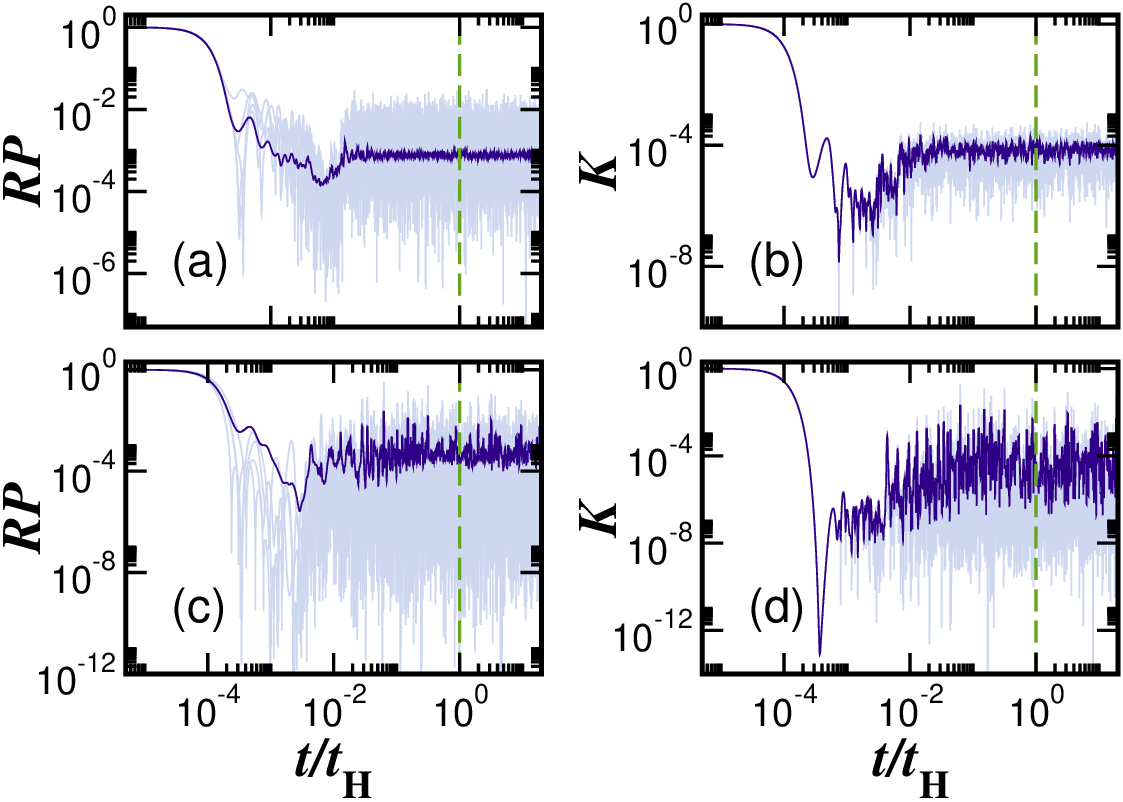}
\caption{Full-time dynamics of $\text{RP}(t)$ and $K(t)$ generated by the $XXZ$ (a, b) and $XX$ (c, d) models. The light-colored curves in panels (a, c) are for individual samples of $\text{RP}(t)$ obtained for five different initial states, meanwhile the dark-colored ones correspond to an average over $200$ initial states, together with an additional moving average performed to further smooth the curves. For $K(t)$ in panels (b, d) the light-colored curves represent the dynamics from a single initial state and the dark-colored curves correspond to moving averages over the same sample. System size is $L=16$.}
\label{fig:dyn2}
\end{figure}

Figure~\ref{fig:dyn2} shows the whole dynamics for the clean $XXZ$ and $XX$ models, but considering only the return probability [Figs.~\ref{fig:dyn2}(a) and~\ref{fig:dyn2}(c)] and spectral form factor [Figs.~\ref{fig:dyn2}(b) and~\ref{fig:dyn2}(d)]. In Figs.~\ref{fig:dyn2}(a) and~\ref{fig:dyn2}(c) the dark-blue curves are averages over $200$ initial states together with an additional moving average to reduce fluctuations. Light-blue curves are dynamics for five individual initial states with energies closest to zero. In Figs.~\ref{fig:dyn2}(b) and~\ref{fig:dyn2}(d) the dark curves are moving averages over a single sample. While the light-colored curves are the rough data. Although interesting transient behavior is observed in Figs.~\ref{fig:dyn} and~\ref{fig:dyn2}, it is worth to emphasize that our main analysis of the distribution functions will be carried out for times beyond the Heisenberg time, $t_\text{H}$. However, we eventually will provide examples of distribution functions for different timescales. The vertical dashed line in all panels of both figures marks the point where $t/t_\text{H}=1$. 

\section{Central Limit Theorem and distribution functions}\label{sec:CLT}
Equation~\eqref{eq:rp} for the return probability can be rewritten as
\begin{equation}\label{eq:spe}
\text{RP}(t)= \left|\sum _{\alpha =1}^{\cal{N}}X_{\alpha}(t) - i \sum _{\alpha =1}^{\cal{N}}Y_{\alpha}(t)\right|^2,
\end{equation}
where $X_\alpha(t)=|c^0_\alpha|^{2}\cos(E_\alpha t)$ and $Y_\alpha(t)=|c^0_\alpha|^{2}\sin(E_\alpha t)$. 
At this point we remind that the classical central limit theorem roughly states that if $X_\alpha(t)$ or $Y_\alpha(t)$ are independent and identically distributed (i.i.d.) random variables with finite second moment, then the distribution of their sum is well approximated by a Gaussian shape. Thus, if we assume that $X_\alpha(t)$ and $Y_\alpha(t)$ behave as i.i.d. in a time window $[t_i,t_f]$ with initial time $t_i$ and final time $t_f$, then Eq.~\eqref{eq:spe} can be considered as the squared norm of a complex random variable with Gaussian real and imaginary parts. In such a case the distribution of $\text{RP}(t)$ should be exponential~\cite{Siddiqui1962},
\begin{equation}\label{eq:RPexp}
P(\text{RP}) = \lambda e^{-\lambda \text{RP}}, \hspace{0.5cm} 
\text{RP}\geq 0, \hspace{0.5cm} \lambda > 0.
\end{equation}
Considering the terms appearing in Eq.~\eqref{eq:spe}, it is clear that if both, initial state components $c^0_\alpha$ and energy eigenvalues $E_\alpha$ behave simultaneously as i.i.d., then the CLT should hold. It is important to mention that in this work (see Appendix~\ref{sec:appA}), we show numerical evidence that a less restrictive condition for the fulfillment of the CLT could be that of uncorrelated random variables.  Assuming that $X_\alpha(t)$ and $Y_\alpha(t)$ remain each one identically distributed, three possible scenarios where the CLT does not hold are when initial state components are autocorrelated, energy eigenvalues are autocorrelated or both of them are correlated, each of these cases mean that $X_\alpha(t)$ and $Y_\alpha(t)$ are not independent. Note, however, that even for weakly enough correlated variables the CLT could remain true~\cite{Hilhorst2009}. 

In the case of $K(t)$ the exponential distribution should be achieved when correlations between energy eigenvalues are absent, just see the expression defining $K(t)$ in Eq.~\eqref{eq:sff}.  The scenario for the expectation value of generic observables is more involved. From its time evolution 
\begin{equation}\label{eq:go}
\left\langle{\cal{O}}(t)\right\rangle= \sum_{\alpha,\beta}c_\alpha^0c_\beta^0{\cal{O}}_{\alpha\beta}e^{-i(E_\alpha-E_\beta)t},   
\end{equation}
with ${\cal{O}}_{\alpha\beta}=\left\langle \psi_\alpha|{\cal{O}}|\psi_\beta\right\rangle$, we see that matrix elements in energy representation of the observable are also included. The simplest case arises with the assumption that all terms inside the summation in Eq.~\eqref{eq:go} behave as i.i.d. random variables, then by the CLT we should expect the distribution to be Gaussian.  We note that having a Gaussian distribution for the late-time values of a generic observable could be an indication that the matrix elements of the observable in energy representation are u.i.d. by themselves. However, if the distribution is not Gaussian, then we cannot ensure that the matrix elements are correlated, this because the presence of energy eigenvalues and eigenstates inside the summation that could also be correlated. Although interesting, we leave the analysis of those fine details for a future work. 

At this point it is fair to mention that Aurich and Steiner conjetured in Ref.~\cite{Aurich1999} that given an extended initial state, the distribution function of a normalized version of the return amplitude $S(t)$ for a chaotic quantum system should be universally described by Rayleigh's law,
\begin{equation}\label{eq:RL}
P(S) = \frac{\pi}{2}S\exp\left(-\frac{\pi}{4}S^{2}\right), \hspace{1cm} S\geq 0.
\end{equation} 
This conjecture was established in the context of 2D and 3D quantum billiards. It is straightforward to show that the square root of an exponential distribution, like the one for the return probability, results in a Rayleigh distribution just as Eq.~\eqref{eq:RL}, this is consistent with Eq.~\eqref{eq:RPexp} which was derived using only basics from probability theory.  Also fair is to note that in the context of RMT, Kunz arrived to the rigorous result that the pdf of both, $K(t)$ and $\text{RP}(t)$ are exponential, independently of the initial state in the later case~\cite{Kunz1999,Kunz2002}.  
\section{Results}\label{Sec:Results}
Now we present and describe our results on the pdfs, starting with the return probability $\text{RP}(t)$, to which we devote majority of this paper. Results about level statistics are also presented, but of course they are no new in the existing literature. As explained before, we use them only as a reference.
\subsection{Return probability and spectral form factor}\label{SubSec. III.A}
\begin{figure}[h]
\centering
\includegraphics[scale=0.345]{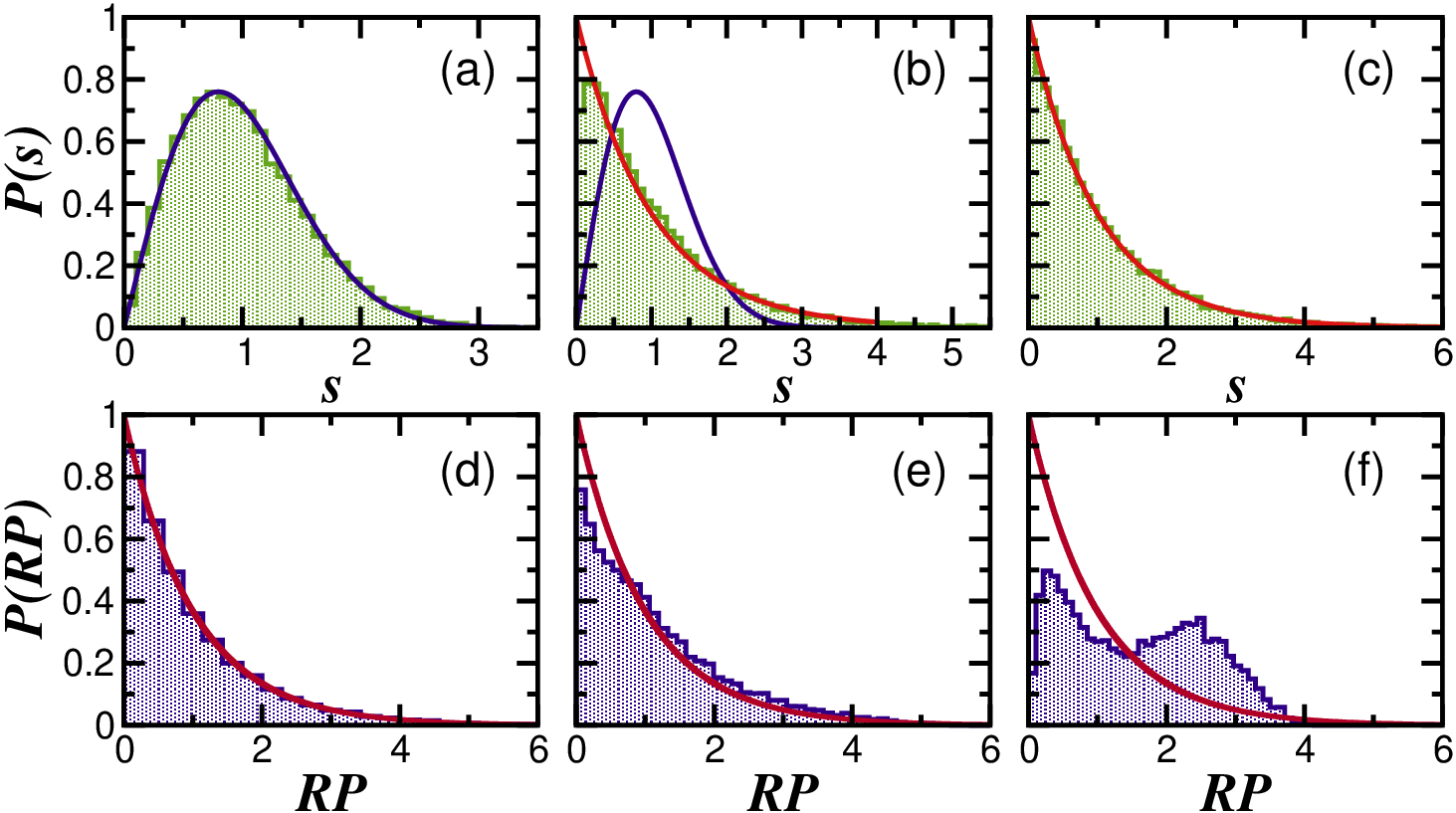}
\caption{Level spacing distribution $P(s)$ (upper panels) and distribution of the late-time values of $\text{RP}(t)$ (lower panels) for the disordered model. The time intervals are $t/t_\text{H}\in(11.38,14.93)$, $t/t_\text{H}\in(29.25,38.38)$, and $t/t_\text{H}\in(33.34,43.74)$ for the disorder strengths $h=0.5$ (a), $h=3.75$ (b) and $h=6.0$ (c), respectively. Solid lines in the upper panels depict the theoretical values for $P(s)$, Eq.~\eqref{eq:P} [color red in panels (b) and (c)] and Eq.~\eqref{eq:GOE} [color blue in panels (a) and (b)]. Solid curves in the lower panels correspond to Eq.~\eqref{eq:RPexp} with mean 1. System size is $L=18$.}
\label{fig:dsp}
\end{figure}

Figure~\ref{fig:dsp} depicts $P(s)$ and $P(\text{RP})$ for the disordered model and fixed number of spins $L=18$. The upper panels $P(s)$ show the behavior already known in the vicinity of the MBL transition~\cite{Serbyn2016}. For small disorder strength, $h=0.5$ [Fig.~\ref{fig:dsp} (a)], $P(s)$ has the shape described by Eq.~\eqref{eq:GOE}. At intermediate disorder strength, $h=3.75$, the shape of $P(s)$ is neither GOE-like nor Poisson-like, although closer to the Poisson shape than to GOE. Once at the localized phase, $h=6.0$, $P(s)$ gets the shape predicted by Eq.~\eqref{eq:P}.

Distributions of $\text{RP}(t)$ with $t/t_\text{H}>1$ are shown in the lower panels of Fig.~\ref{fig:dsp}, specific time intervals are indicated in the captions. An initial state with energy $E_0$ closest to zero is chosen and a single disorder realization is considered. Note that in the whole work we have normalized the values of $\text{RP}(t)$ with their standard deviation in the respective time interval. Figure~\ref{fig:dsp} (d) confirms the anticipated in Sec.~\ref{sec:CLT}, for $h=0.5$ the distribution of RP (histogram in color blue) conforms with the exponential distribution given by Eq.~\eqref{eq:RPexp} with $\lambda=1$. Then we could assume that $c_\alpha^0$, $E_\alpha$, $X_{\alpha}(t)$ and $Y_{\alpha}(t)$ behave as uncorrelated random variables and therefore the universal exponential pdf is achieved (see discussion in Appendix~\ref{sec:appA}, where evidence is given in the sense that the assumption is actually a fact).The increase of disorder strength $h$ is accompanied by the lack of correspondence between the exponential and the distributions of RP, which is clearly seen in Figs.~\ref{fig:dsp}(e) and~\ref{fig:dsp}(f) for $h=3.75$ and $h=6.0$, respectively. This is a relevant observation because the distribution $P(\text{RP})$ could detect the MBL transition, and remarkably from a single sample of the dynamics. However, we emphasize that detection of the MBL transition is not the one of the aims of this work, which instead is the establishment of a Gaussian scenario for the equilibrium state of interacting quantum systems.

The shape of $P(\text{RP})$ different from the exponential is a signature of autocorrelations, be in the energy eigenvalues or between initial state components. For this last statement, we can apply the negation of the CLT, if the distribution of the sums involved in Eq.~\eqref{eq:spe} does not tend to a Gaussian, then the random variables are not statistical independent, they are correlated. Correlations between eigenstates components have been shown to exist around the MBL transition; see, for instance, Refs.~\cite{Torres2015,Torres2017} and more recently Ref.~\cite{Tikhonov2021}. In fact, the slow dynamics produced by those correlations have been observed experimentally~\cite{Luschen2017}. 

\begin{figure}[H]
\centering
\includegraphics[scale=0.37]{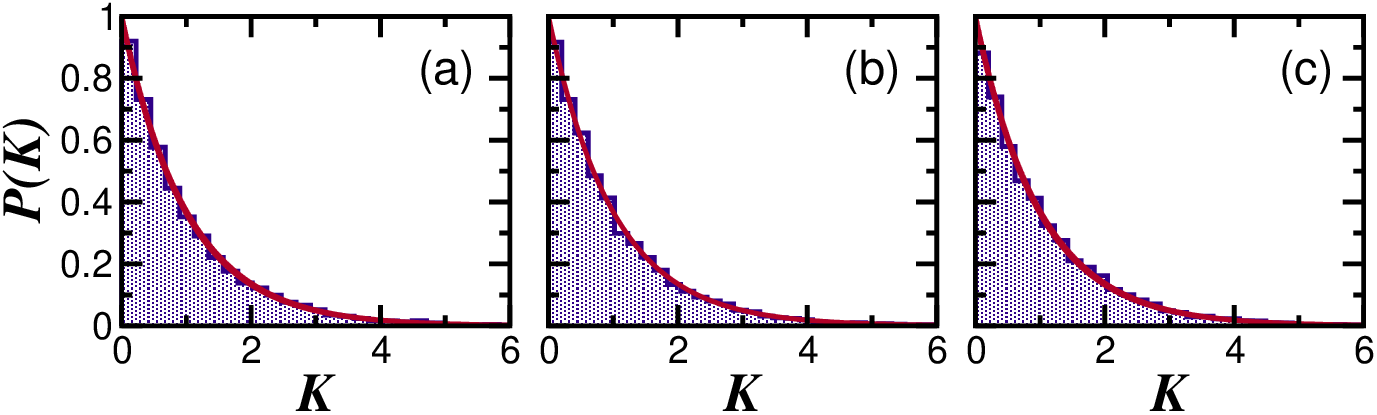} 
\caption{Distribution of the late-time values of $K(t)$ for the same time window as in Fig.~\ref{fig:dsp} for the disordered model with $h=0.5$ (a), $h=3.75$ (b), and $h=6.0$ (c). Solid curves are given by $P(K)=\exp(-K)$. System size is $L=18$.}
\label{fig:dsff}
\end{figure}

\begin{figure}[t]
\centering
\includegraphics[scale=0.36]{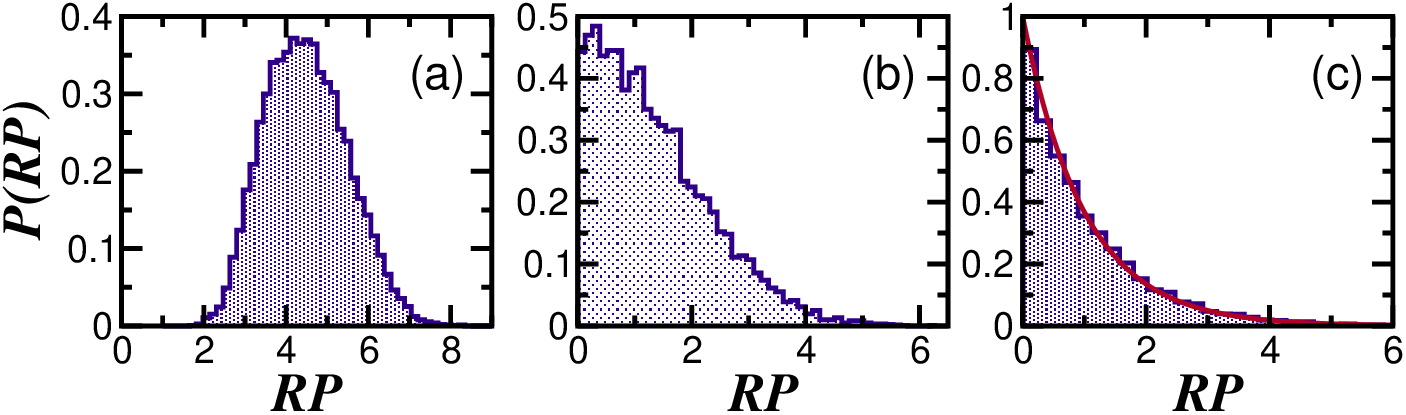} 
\caption{Distribution of the late-time values of $\text{RP}(t)$ under the disordered model in the ergodic phase with $h=0.5$. The considered time window is $t/t_\text{H}\in(11.38, 14.93)$. The initial state is a random one with correlated components $c_\alpha^0$. Distributions for different degrees of correlation are depicted: strong $q=0.5$ (a), intermediate $q=0.33$ (b) and weak $q=0.2$ (c). Red solid curve in panel (c) corresponds to $P(\text{RP})=\exp(-\text{RP})$. System size is $L=18$.}
\label{fig:xxzc}
\end{figure}

However, up to our knowledge, information about autocorrelations of energy levels does not exist. To fill out this gap, we recur to the spectral form factor and analyze its distribution for $t/t_\text{H}>1$. Figure~\ref{fig:dsff} is conclusive, it shows that for any disorder strength in the MBL model the distribution of $K(t)$ has an exponential shape. This exponential shape can be explained analogously to the case of $\text{RP}(t)$, that is, in terms of the CLT. As suggested by the results depicted Fig.~\ref{fig:AXXZh}.1 (a) of Appendix~\ref{sec:appA}, the energy eigenvalues remain uncorrelated for any disorder strength. Then the sole cause for the unconform shapes of $P(\text{RP})$ in Figs.~\ref{fig:dsp}(e) and~\ref{fig:dsp}(f) should be the autocorrelations between initial state components $c_\alpha^0$. To reinforce our statement about autocorrelations between initial state components, we present in Fig.~\ref{fig:xxzc} the distribution $P(\text{RP})$ departing from a random initial state with tuned degree of autocorrelations and evolved by the disordered model in the chaotic region, with fixed $h=0.5$. There the components $c_\alpha^0$ of the initial state are random numbers from a uniform distribution and with autocorrelation degree $q$, generated by the recipe given in Ref.~\cite{Thomas1993} and recently used to show the effects of correlated disorder in the region around the MBL transition in Ref.~\cite{Vallejo2022}. Figure~\ref{fig:xxzc} shows that as the degree of autocorrelations moves from a regime of strong autocorrelations (a), passing trough intermediate (b) and finally arriving to weak autocorrelations (c), the exponential distribution of $P(\text{RP})$ is eventually achieved. The autocorrelations of initial state components are confirmed through Fig~\ref{fig:AXXZh}.1 of Appendix~\ref{sec:appA}, there it is shown that while, as already said, energy eigenvalues remain uncorrelated for any disorder strength, the initial state components get autocorrelated once the disorder strength is strong enough, say $h=3.75$. Thus apparently inducing autocorrelations in $X_{\alpha}(t)$ and $Y_{\alpha}(t)$, as suggested also by the results shown in Figs.~\ref{fig:AXXZh}.1(c) and~\ref{fig:AXXZh}.1(d), respectively.  

\subsubsection{XX and XXZ models}
\begin{figure}[t]
\centering
\includegraphics[scale=0.51]{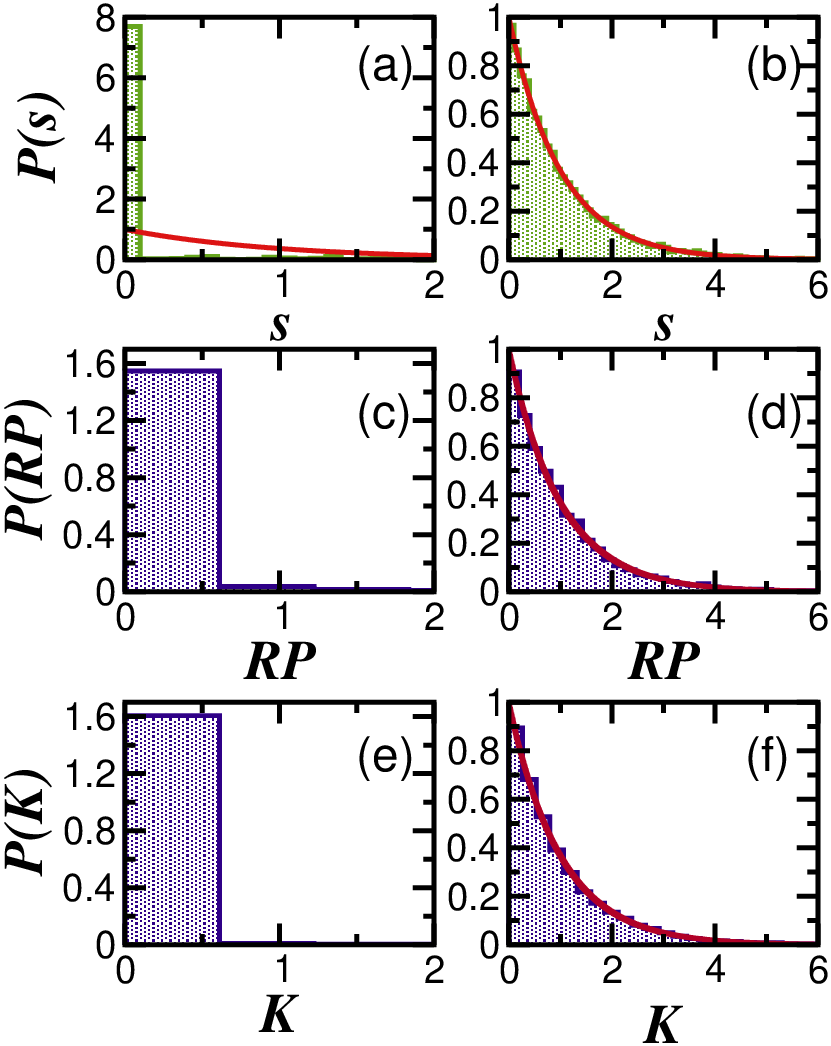}
\caption{Distribution of spacings $P(s)$ (upper panels), distribution of the late-time values of $\text{RP}(t)$ (middle panels) and $K(t)$ (lower panels). Left column contains results for the $XX$ model, meanwhile in the right columns the results are for the $XXZ$ model. The considered time windows are $t/t_\text{H}\in(9.092,11.928)$ for the $XX$ model and $t/t_\text{H}\in(9.859,12.935)$ for the $XXZ$ model. $L=18$.}
\label{fig:dxxxxz}
\end{figure}

As already discussed, autocorrelations between energy levels could block the fulfillment of the CLT. Those correlations can be caused, for instance, by degeneracies in the energy spectrum. A suitable model to show this is the $XX$ model [$\Delta=0$ and $h=0$ in Eq.~\eqref{eq:Ham}], which energy spectrum contains a high amount of degeneracies~\cite{Zangara2013}. Figure~\ref{fig:dxxxxz} (a) showing $P(s)$ for the $XX$ model confirms the degeneracies with the Shnirelman's peak as a witness~\cite{Chirikov1995}. Consequently, the distribution not only of $\text{RP}(t)$ but also of $K(t)$ should not be exponential, which is confirmed by Figs.~\ref{fig:dxxxxz}(c) and~\ref{fig:dxxxxz}(e) for $P(\text{RP})$ and $P(K)$, respectively. 

We complement our analysis with the study of the time evolution generated by the $XXZ$ model [$\Delta=1$ and $h=0$ in Eq.~\eqref{eq:Ham}]. The Ising interactions present in the integrable $XXZ$ model break some symmetries of the $XX$ model and thus removes degeneracies in the energy spectrum. This is reflected into $P(s)$, Fig.~\ref{fig:dxxxxz}(b), which displays Poisson-like statistics. The distribution of the return probability, $P(\text{RP})$, is exponential meaning that energy eigenvalues and initial state components behave as uncorrelated random variables. 

Figure~\ref{fig:dxxxxz} for $P(K)$ confirms absence of autocorrelations between eigenvalues. To confirm the last same for initial state components, we employed the so-called inverse participation ratio, $\text{IPR}=\sum _{\alpha=1}^{\cal{N}}|\braket{\psi_{\alpha}|\Psi(0)}|^{4}\propto {\cal{N}}^{-D_2}$, with $0\leq D_2\leq 1$ known as the correlation dimension that measures the degree of correlations between eigenstate components~\cite{Wegner1980}. Two extreme cases $D_2=0$ and $D_2=1$ mean correlations and absence of correlations, respectively. Our scaling analysis of IPR with dimension $\cal{N}$ (not shown) for the initial state $\ket{\Psi(0)}$ used in Fig.~\ref{fig:dxxxxz} (d) lead us to $D_2\approx 0.75$, this value suggests weak correlations between initial state components $c_\alpha^0$, at least weak enough to have the CLT still valid.

Our results for the integrable $XXZ$ model reaffirm the idea that the universal late-time exponential distribution is achieved once the conditions for the CLT fulfillment are provided, that is, simultaneous absence or weak autocorrelations in the initial state components and in the energy spectrum of the Hamiltonian generating the time evolution.

\subsubsection{Timescales}\label{SubSec. III.C}
\begin{figure}[t]
\centering
\includegraphics[scale=0.38]{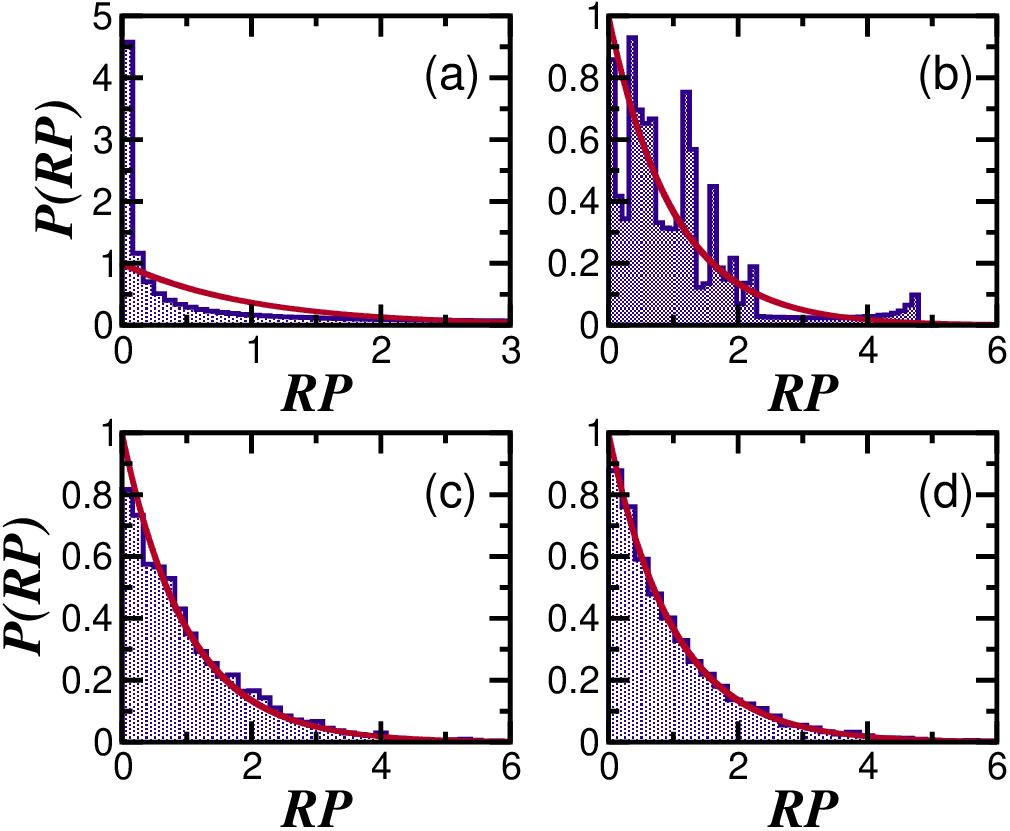} 
\caption{Distribution $P(\text{RP})$ of the return probability for the disordered model in the ergodic phase with $h=0.5$ in different time windows. Initial decay, $t/t_\text{H}\in(0.000034,0.000089)$ (a), power-law decay, $t/t_\text{H}\in(0.00034,0.00089)$ (b), around the time where the minimum of the correlation hole (Thouless time) is located, $t/t_\text{H}\in(0.0091,0.0228)$ (c), and saturation, $t/t_\text{H}\in(11.38,14.93)$ (d). Solid curves are given by $P(\text{RP}) = \exp(-\text{RP})$. System size is $L=18$.}
\label{fig:tw}
\end{figure}

Up to now, the analysis has been carried out using timescales greater than the Heisenberg time $t_\text{H}$, an interesting question is if the universal distribution could appear for smaller timescales. Figure~\ref{fig:tw} displays the distribution of RP for the disordered model in the ergodic phase with $h=0.5$ and for different time intervals, at very short times [Fig.~\ref{fig:tw}(a)], power-law decay [Fig.~\ref{fig:tw}(b)], around the correlation hole [Fig.~\ref{fig:tw}(c)], and saturation [Fig.~\ref{fig:tw}(d)]. The reader is referred to Fig.~\ref{fig:dsp}(a) and discussion in Sec.~\ref{sec:dynamics} to remind all those timescales.

Once the initial state begins the transition to other basis states, as time increases, the distribution $P(\text{RP})$ moves gradually to an exponential one. At short times the return probability is very similar in values and a peaked distribution is obtained, Fig.~\ref{fig:tw}(a). The exponential shape is completely achieved when the time is large enough like in Fig.~\ref{fig:tw}(d); however, note that for times when the correlation hole appears (Thouless time) [Fig.~\ref{fig:tw}(c)] the distribution of RP also looks like an exponential but not quite good as in Fig.~\ref{fig:tw}(d). Then we infer that to observe the universal exponential distribution one should wait for times beyond the Heisenberg time, $t_\text{H}$. Now we have the opportunity to mention that the exponential distribution of $\text{RP}(t)$ in a time interval coincides with the exponential observed in Ref.~\cite{TorresHerrera2020b} also at large times, but there fixing a single time $t_0$ and computing the pdf over different disorder realizations and initial states. This is consistent with the fact that to study ergodicity one should do it once at equilibrium.

\subsection{Spin autocorrelation function and connected spin-spin correlation function}\label{SubSec. III.B}

\begin{figure}[t]
\centering
\includegraphics[scale=0.3645]{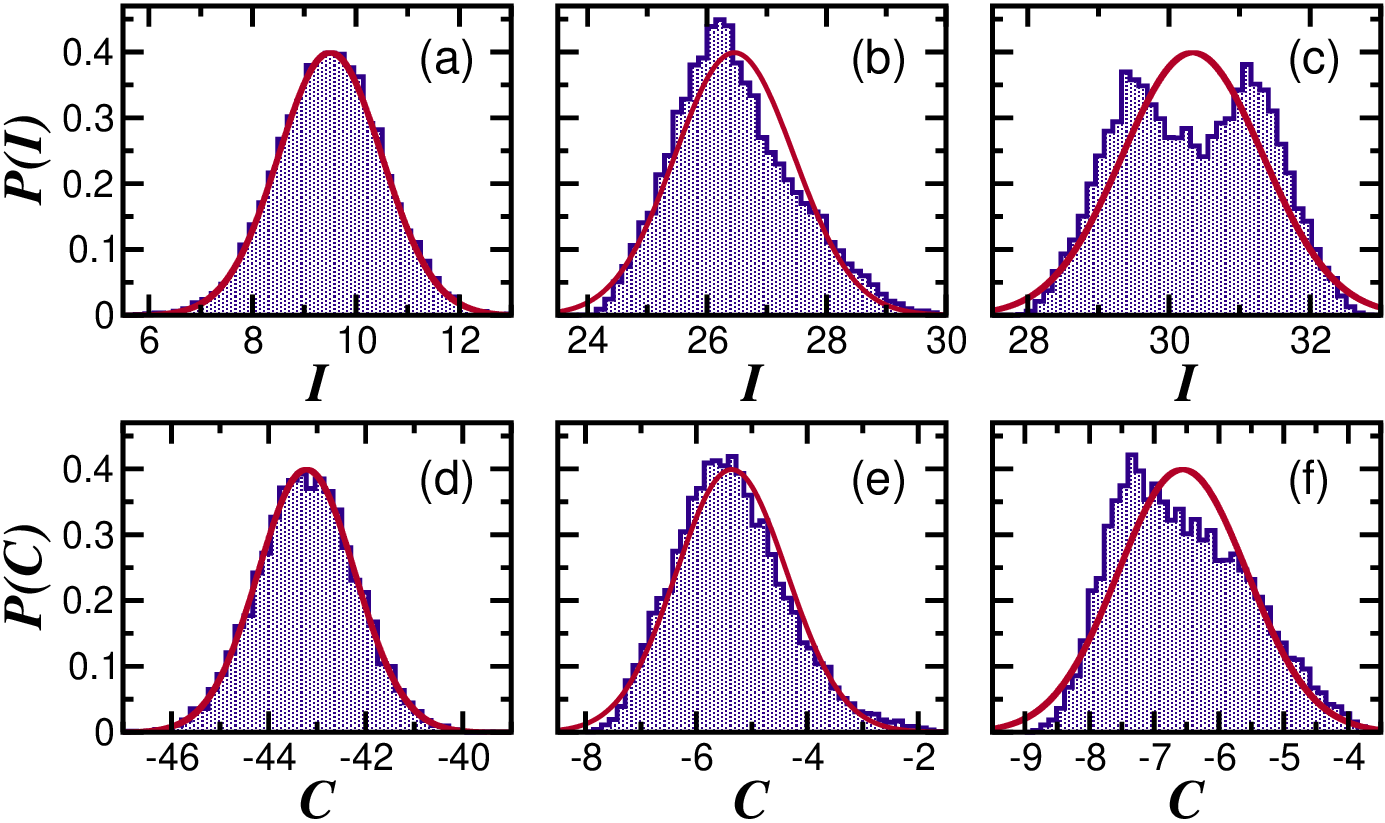} 
\caption{Late-time probability distribution functions corresponding to the spin autocorrelation function (upper panels) and the connected spin-spin correlation function (lower panels) evolved under the disordered model. The considered time windows are the same as in Fig.~\ref{fig:dsp}. Disorder strengths are $h=0.5$ (a, d), $h=3.75$ (b, e), and $h=6.0$ (c, f). Solid curves are Gaussian functions, $P(x)=(\sqrt{2\pi})^{-1/2}e^{-(x-\mu)^2/2}$, with $\mu$ the mean of the corresponding observable. System size is $L=18$.}
\label{fig:obs}
\end{figure}

In this section we focus on the analysis of the pdf of two dynamical observables of great interest in quantum simulations with cold atoms and ion traps, the spin autocorrelation function $I(t)$ and the connected spin-spin correlation function $C(t)$ given by Eqs. (\ref{eqn:CSSCF}) and (\ref{eqn:SAF}), respectively. We see in the upper panels of Fig.~\ref{fig:obs} the pdf for the spin autocorrelation function, $P[I(t/t_\text{H}>1)]$, meanwhile in the lower panels of Fig.~\ref{fig:obs} we have the pdf for the connected spin-spin correlation function, $P[C(t/t_\text{H}>1)]$. Both quantities are evolved by the disordered model with different disorder strengths. As predicted in Sec.~\ref{sec:CLT}, when $h=0.5$ the late-time pdf are Gaussian for both observables [Figs.~\ref{fig:obs}(a) and~\ref{fig:obs}(d)]. 
Once the system is outside the ergodic phase, the pdf for both observables do not conform with the Gaussian shape. Comparing Figs.~\ref{fig:obs}(b) and~\ref{fig:obs}(e), for $h=3.75$, we see that $P(I)$ is more sensitive to the change in disorder than $P(C)$. Finally, when the disorder strength is $h=6.0$, the lack of correspondence between the pdf of both observables and the Gaussian is even more evident. Again, the shape of the pdf could be used as a probe of the MBL transition. Interestingly, our results in the ergodic side of the MBL are in line with results for noninteracting systems in Refs.~\cite{Venuti2010,Venuti2013}.

\section{Conclusions}\label{Sec:Conclusions}
We analyzed the late-time probability distribution functions of dynamical quantities of interest for experimental platforms. We argued that the pdfs are universal in the sense that the unique requirement is the CLT to hold. This needs uncorrelated initial state components and energy spectrum. For chaotic quantum systems these conditions are fulfilled, but also for integrable systems like the $XXZ$ model. Strong degeneracies in the energy spectrum, like in the $XX$ model, prevent the CLT to hold, as well as correlated initial state components, like in the disordered model for intermediate and strong disorder strengths. For generic observables, an additional condition should be fulfilled, uncorrelated elements in energy representation. This happens for the two quantities studied in this work, the spin autocorrelation function and the connected spin-spin correlation function.  No averages are needed, our analysis was based on a single sample of the dynamics. Details about autocorrelations between initial state components, energy levels and elements of observables deserve a deeper analysis, but this is leaved for a future work.  We expect our results to motivate further studies of late-time pdfs in several quantum systems. 

\begin{acknowledgments}
I.V.-F. and E.J.T.-H. are grateful to LNS-BUAP for their supercomputing facility. E.J.T.-H. is also grateful for financial support from VIEP-BUAP, Project No. 00270-2022 and CONAHCYT under Project Ciencia de Frontera No. CF-2023-I-1748.
\end{acknowledgments}

\appendix
\renewcommand\thefigure{\thesection.\arabic{figure}} 

\setcounter{figure}{0}
\section{Autocorrelations}\label{sec:appA}
To test the lack or presence of autocorrelations between eigenvalues $E_\alpha$, initial state components $c_\alpha^0$ and the variables $X_\alpha(t)$ and $Y_\alpha(t)$ defined in the context of Eq.~\eqref{eq:spe}, we employ the so-called autocorrelation function, given by
\begin{equation}\label{eq:ACF}
R_x(k) = \frac{1}{({\cal{N}}-k)\sigma^2} \sum_{\alpha=1}^{{\cal{N}}-k} (x_\alpha - \mu)(x_{\alpha+k} - \mu).
\end{equation}
This function measures the linear correlation between the sequence of values $x_\alpha$ and a copy of it at lag $k$, that is, $x_{\alpha+k}$. In Eq.~\eqref{eq:ACF}, $\mu=\left\langle x_\alpha\right\rangle$ and $\sigma^2=\left\langle x_\alpha^2\right\rangle-\left\langle x_\alpha\right\rangle^2$ are the mean and variance, respectively, of $x_\alpha$. Having $R_x(k)=0$ for all lag $k$ means absence of autocorrelations. Meanwhile, $R_x(k)=1$ for some $k$ indicates maximum level of autocorrelations, which by definition is always the case for $k=0$. In the following we avoid $k=0$ and restrict ourselves to $k=1,\,\dots,\,{\cal{N}}/2$, this because of in our case $k$ larger than ${\cal{N}}/2$ do not provide additional information.

\begin{figure}[t]\label{fig:AXXZh}
    \centering
    \includegraphics[width=\columnwidth]{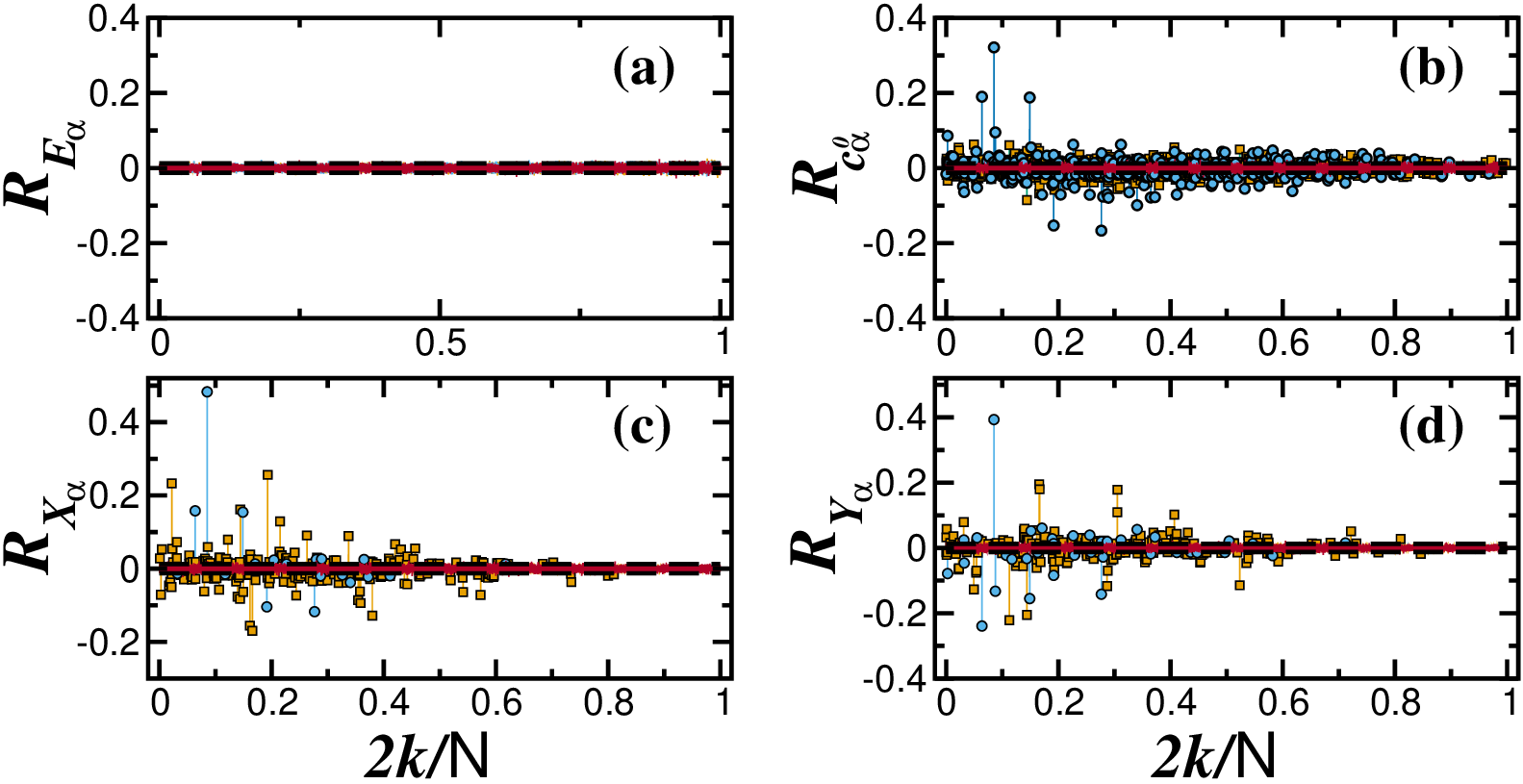}
    \caption{Autocorrelation function $R$ for $E_\alpha$ (a), 
 $c_\alpha^0$ (b),  $X_\alpha(t)$ (c), and $Y_\alpha(t)$ (d). Results are for the MBL model with $h=0.50$ (red), $h=3.75$ (orange) and $h=6.00$ (blue). For panels (c, d), we consider $t/t_H\approx4.553$ ($h=0.5$), $t/t_H\approx11.701$ ($h=3.75$), and $t/t_H\approx13.334$ ($h=6.0$). Dashed lines determine a confidence interval indicated in the main text. System size is $L=18$.} 
\end{figure}

Our results for the autocorrelation function $R_x(k)$ are shown in Fig.~\ref{fig:AXXZh}.1. We consider the MBL model with $h=0.5$ (red), $h=3.75$ (orange, squares), and $h=6.0$ (blue, circles). The system size if fixed to $L=18$. Note that we normalize $k$ with ${\cal{N}}/2$. All panels of Fig.~\ref{fig:AXXZh}.1 include a pair of horizontal dashed lines delimiting a standard confidence interval of 95\% computed as $\pm1.96/\sqrt{\cal{N}}$, thus $R_x(k)$ staying inside this interval for all $k$ suggests lack of autocorrelations.   

Figure~\ref{fig:AXXZh}.1(a) confirms what was already anticipated as a naive assumption in the discussion around Fig.~\ref{fig:dsff} for the late-time distribution function of the spectral form factor $K(t)$, energy eigenvalues remain uncorrelated for any value of the disorder strength $h$. This is witnessed by the behavior of $R_{E_\alpha}(k)$, which is inside of the confidence interval for all lag $k$. 

Figure~\ref{fig:AXXZh}.1(b) for $R_{c_\alpha^0}(k)$ is quite interesting, for $h=0.5$ its value for any lag $k$ lies inside the confidence interval, suggesting that the initial state components $c_\alpha^0$ are not autocorrelated. In contrast, since $R_{c_\alpha^0}(k)$ lies notably outside of the confidence interval for some lags, the image for $h=3.75$ and $h=6.0$ is that of autocorrelated components. In more detail, we observe that for $h=6.0$ some values of $R_{c_\alpha^0}(k)$ are larger than for $h=3.75$. 
Apparently the autocorrelations in $c_\alpha^0$ are the origin of autocorrelations in $X_\alpha(t)$ and $Y_\alpha(t)$, as shown in Figs.~\ref{fig:AXXZh}.1(c) and~\ref{fig:AXXZh}.1(d). The behavior of $R_{X_\alpha}(k)$ and $R_{Y_\alpha}(k)$ is similar to the one just described for $R_{c_\alpha^0}(k)$, having values larger than the confidence interval only for $h=3.75$ and $h=6.0$, while for $h=0.5$ the values are inside the confidence interval for any lag $k$. These observations are fully consistent with Figs.~\ref{fig:dsp} and~\ref{fig:xxzc} for the return probability and the discussion around them. They confirm the conditions for the fulfillment of the CLT in the context of Eq.~\eqref{eq:spe}.   

Our results with smaller system sizes $L=8-16$, not shown here, are consistent with the ones for $L=18$. Of course, the smaller the system size the wider the confidence interval of 95\%.  We leave a deeper analysis on autocorrelations for a future work.

Let us make a final comment. Certainly, it is known that eigenvalues of the MBL model in the thermalizing regime, say with $h=0.5$, show level repulsion just as eigenvalues of full random matrices from a Gaussian orthogonal ensemble~\cite{Serbyn2016}. This is what in the context of RMT is known as a correlated energy spectrum. Then, our claims based on the behavior of $R_{E_\alpha}(k)$ depicted in Fig.~\ref{fig:AXXZh}.1 could appear as intriguing and surprising. However, one should remind that the autocorrelation function $R_x(k)$ measures only linear autocorrelations, and the lack of those kind of autocorrelations appears as enough to achieve fulfillment of the CLT and consequently universal probability distribution functions.

\setcounter{figure}{0}
\section{System-size dependence}\label{sec:appB}
\begin{figure}[t!]\label{fig:SSSObs}
    \centering
    \includegraphics[width=\columnwidth]{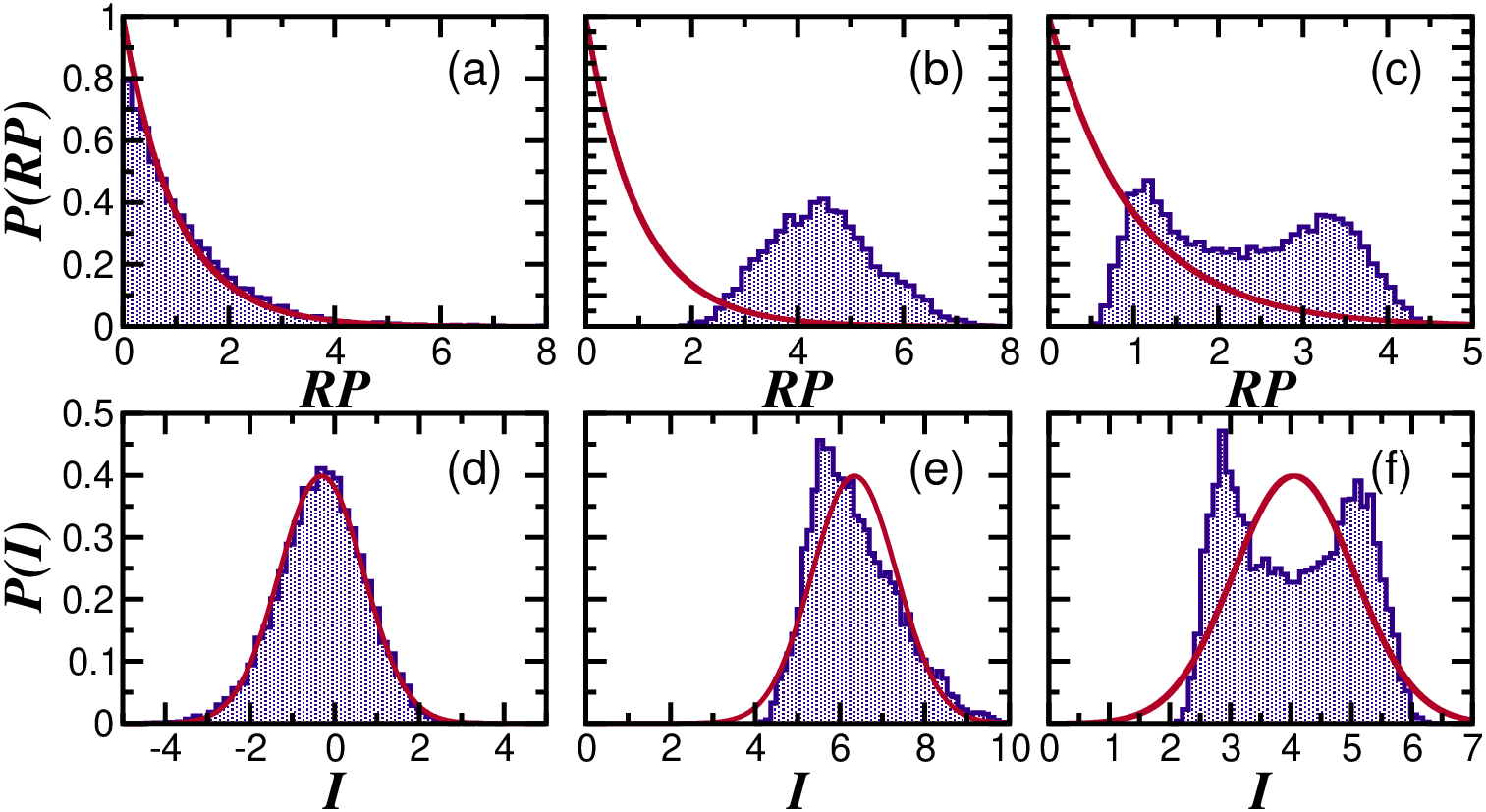}
    \caption{Probability distribution function of the return probability (upper panels) and imbalance (lower panels) for the MBL model with $L=6$. $h=0.5$ in panels (a) and (d), $h=3.75$ in panels (b) and (e), $h=6.0$ in panels (c) and (f). Time intervals as in Figs.~\ref{fig:dsp} and~\ref{fig:obs}. Solid curves in the upper panels are Eq.~\eqref{eq:RPexp} with $\lambda=1$, while in the lower panels they are Gaussian fittings.}
\end{figure}

Our results in the main text where carried out for a fixed system size, $L=18$, which certainly is not that big, but large enough for standard exact diagonalization. Here we explore a much smaller system size, say $L=6$. Figure~\ref{fig:SSSObs}.1 shows the pdf of the return probability $\text{RP}(t)$ and the spin autocorrelation function, $I(t)$. The observed images are similar to the lower panels of Fig.~\ref{fig:dsp} for $\text{RP}(t)$ and upper panels of Fig.~\ref{fig:obs} for $I(t)$. Exponential [Fig.~\ref{fig:SSSObs}.1 (a)] and  Gaussian [Fig.~\ref{fig:SSSObs}.1 (d)] pdfs for $\text{RP}(t)$ and $I(t)$ when the system is in the ergodic phase with $h=0.5$. For stronger disorder strength, $h=6.0$, the analysis leads to pdfs that do not conform with exponential [Fig.~\ref{fig:SSSObs}.1 (c)] or Gaussian [Fig.~\ref{fig:SSSObs}.1 (f)]. The picture for $h=3.75$ goes in line with the one for $h=6.0$. 

The considered system size, $L=6$, is the smallest one we can take into account to avoid what apparently are finite size effects. Our results for $L=4$ (not shown here) with half-filling display pdfs that do not conform for any disorder strength with the Gaussian scenario that we have established for interacting quantum systems. 
Something comparable is expected to happen for other models like the clean $XXZ$ model.

\setcounter{figure}{0}
\section{Disorder realization dependence}\label{sec:appC}
We devote this Appendix to the analysis of the dependence on the disorder realization of our results for the MBL model. It is justified since one reader could think, with certain reason, that independent random disorder realizations will result in different pdfs. Certainly, it is known that the time evolution of physical quantum models depends on the initial state, then in the case of the MBL model, for each disorder realization, if the initial state is not fixed by hand, then a different initial state could be picked up and consequently a different pdf could be obtained for the quantities and observables that we are addressing in this work. As stated in the main text, to study the pdf we choose one initial state with energy closest to zero, this is our fixed criteria. 
\begin{figure}[ht!]\label{fig:Real}
    \centering
    \includegraphics[width=0.805\columnwidth]{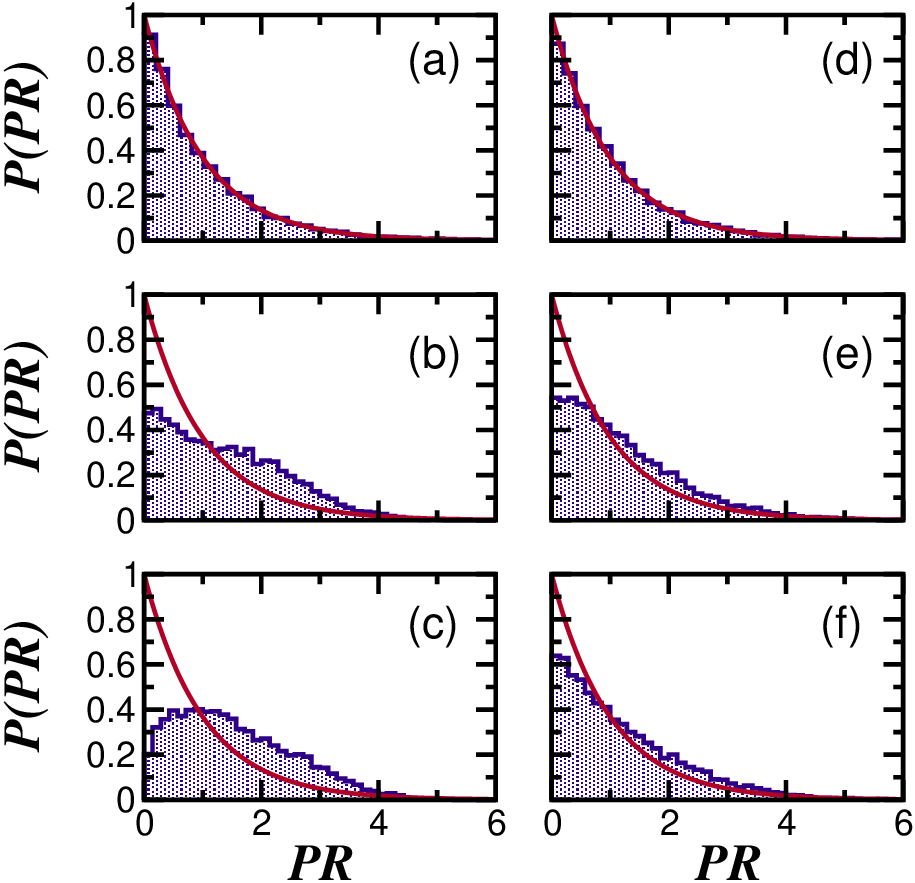}
    \caption{Pdf of the return probability for the MBL model from two different disorder realizations (left and right). $h=0.5$ in panels (a, d), $h=3.75$ in panels (b, e), $h=6.0$ in panels (c, f). Solid curves are the exponential pdf given by Eq.~\eqref{eq:RPexp} with $\lambda=1$. $L=16$.}
\end{figure}
Figure~\ref{fig:Real}.1 shows the pdf of the return probability from two different random realizations of the MBL model and three different disorder strengths. There is seen that the behavior is in general very similar for both realizations (left and right). Specifically, the pdf for $h=0.5$ is the universal exponential distribution, meanwhile for $h=3.75$ and $h = 6.0$ the pdf does not conform with the exponential, although for this two later values the pdf is different for each disorder realization. This can be explained because in the ergodic phase and close to the middle of the energy spectrum, the structure of basis states (one of which is our initial state) is pretty similar and stable, independently of the disorder realization. In contrast, for $h$ around $3.75$ and larger, the distribution of basis states in energy is disperse and the chosen initial state at the middle of the spectrum will have a different structure depending of the disorder realization, but its components will always contain a degree of correlations strong enough to avoid the fullfilment of the central limit theorem.


\begin{thebibliography}{83}%
\makeatletter
\providecommand \@ifxundefined [1]{%
 \@ifx{#1\undefined}
}%
\providecommand \@ifnum [1]{%
 \ifnum #1\expandafter \@firstoftwo
 \else \expandafter \@secondoftwo
 \fi
}%
\providecommand \@ifx [1]{%
 \ifx #1\expandafter \@firstoftwo
 \else \expandafter \@secondoftwo
 \fi
}%
\providecommand \natexlab [1]{#1}%
\providecommand \enquote  [1]{``#1''}%
\providecommand \bibnamefont  [1]{#1}%
\providecommand \bibfnamefont [1]{#1}%
\providecommand \citenamefont [1]{#1}%
\providecommand \href@noop [0]{\@secondoftwo}%
\providecommand \href [0]{\begingroup \@sanitize@url \@href}%
\providecommand \@href[1]{\@@startlink{#1}\@@href}%
\providecommand \@@href[1]{\endgroup#1\@@endlink}%
\providecommand \@sanitize@url [0]{\catcode `\\12\catcode `\$12\catcode
  `\&12\catcode `\#12\catcode `\^12\catcode `\_12\catcode `\%12\relax}%
\providecommand \@@startlink[1]{}%
\providecommand \@@endlink[0]{}%
\providecommand \url  [0]{\begingroup\@sanitize@url \@url }%
\providecommand \@url [1]{\endgroup\@href {#1}{\urlprefix }}%
\providecommand \urlprefix  [0]{URL }%
\providecommand \Eprint [0]{\href }%
\providecommand \doibase [0]{https://doi.org/}%
\providecommand \selectlanguage [0]{\@gobble}%
\providecommand \bibinfo  [0]{\@secondoftwo}%
\providecommand \bibfield  [0]{\@secondoftwo}%
\providecommand \translation [1]{[#1]}%
\providecommand \BibitemOpen [0]{}%
\providecommand \bibitemStop [0]{}%
\providecommand \bibitemNoStop [0]{.\EOS\space}%
\providecommand \EOS [0]{\spacefactor3000\relax}%
\providecommand \BibitemShut  [1]{\csname bibitem#1\endcsname}%
\let\auto@bib@innerbib\@empty
\bibitem [{\citenamefont {Clusel}\ and\ \citenamefont
  {Bertin}(2008)}]{Clusel2008}%
  \BibitemOpen
  \bibfield  {author} {\bibinfo {author} {\bibfnamefont {M.}~\bibnamefont
  {Clusel}}\ and\ \bibinfo {author} {\bibfnamefont {E.}~\bibnamefont
  {Bertin}},\ }\bibfield  {title} {\bibinfo {title} {Global fluctuations in
  physical systems: A subtle interplay between sum and extreme value
  statistics},\ }\href {\doibase 10.1142/S021797920804853X} {\bibfield  {journal} {\bibinfo  {journal} {Int.
  J. Mod. Phys. B}\ }\textbf {\bibinfo {volume} {22}},\ \bibinfo {pages} {3311}
  (\bibinfo {year} {2008})}\BibitemShut {NoStop}%
\bibitem [{\citenamefont {Peres}(1984)}]{Peres1984}%
  \BibitemOpen
  \bibfield  {author} {\bibinfo {author} {\bibfnamefont {A.}~\bibnamefont
  {Peres}},\ }\bibfield  {title} {\bibinfo {title} {Stability of quantum motion
  in chaotic and regular systems},\ }\href {\doibase 10.1103/PhysRevA.30.1610} {\bibfield  {journal}
  {\bibinfo  {journal} {Phys. Rev. A}\ }\textbf {\bibinfo {volume} {30}},\
  \bibinfo {pages} {1610} (\bibinfo {year} {1984})}\BibitemShut {NoStop}%
\bibitem [{\citenamefont {Venuti}\ and\ \citenamefont
  {Zanardi}(2010)}]{Venuti2010}%
  \BibitemOpen
  \bibfield  {author} {\bibinfo {author} {\bibfnamefont {L.~C.}\ \bibnamefont
  {Venuti}}\ and\ \bibinfo {author} {\bibfnamefont {P.}~\bibnamefont
  {Zanardi}},\ }\bibfield  {title} {\bibinfo {title} {Universality in the
  equilibration of quantum systems after a small quench},\ }\href {\doibase 10.1103/PhysRevA.81.032113} {\bibfield  {journal} {\bibinfo  {journal} {Phys. Rev. A}\ }\textbf {\bibinfo
  {volume} {81}},\ \bibinfo {pages} {032113} (\bibinfo {year}
  {2010})}\BibitemShut {NoStop}%
\bibitem [{\citenamefont {Venuti}\ and\ \citenamefont
  {Zanardi}(2013)}]{Venuti2013}%
  \BibitemOpen
  \bibfield  {author} {\bibinfo {author} {\bibfnamefont {L.~C.}\ \bibnamefont
  {Venuti}}\ and\ \bibinfo {author} {\bibfnamefont {P.}~\bibnamefont
  {Zanardi}},\ }\bibfield  {title} {\bibinfo {title} {Gaussian equilibration},\
  }\href {\doibase 10.1103/PhysRevE.87.012106} {\bibfield  {journal} {\bibinfo  {journal} {Phys. Rev. E}\
  }\textbf {\bibinfo {volume} {87}},\ \bibinfo {pages} {012106} (\bibinfo
  {year} {2013})}\BibitemShut {NoStop}%
\bibitem [{\citenamefont {Zangara}\ \emph {et~al.}(2013)\citenamefont
  {Zangara}, \citenamefont {Dente}, \citenamefont {Torres-Herrera},
  \citenamefont {Pastawski}, \citenamefont {Iucci},\ and\ \citenamefont
  {Santos}}]{Zangara2013}%
  \BibitemOpen
  \bibfield  {author} {\bibinfo {author} {\bibfnamefont {P.~R.}\ \bibnamefont
  {Zangara}}, \bibinfo {author} {\bibfnamefont {A.~D.}\ \bibnamefont {Dente}},
  \bibinfo {author} {\bibfnamefont {E.~J.}\ \bibnamefont {Torres-Herrera}},
  \bibinfo {author} {\bibfnamefont {H.~M.}\ \bibnamefont {Pastawski}}, \bibinfo
  {author} {\bibfnamefont {A.}~\bibnamefont {Iucci}},\ and\ \bibinfo {author}
  {\bibfnamefont {L.~F.}\ \bibnamefont {Santos}},\ }\bibfield  {title}
  {\bibinfo {title} {Time fluctuations in isolated quantum systems of
  interacting particles},\ }\href {\doibase 10.1103/PhysRevE.88.032913} {\bibfield  {journal} {\bibinfo
  {journal} {Phys. Rev. E}\ }\textbf {\bibinfo {volume} {88}},\ \bibinfo
  {pages} {032913} (\bibinfo {year} {2013})}\BibitemShut {NoStop}%
\bibitem [{\citenamefont {Gring}\ \emph {et~al.}(2012)\citenamefont {Gring},
  \citenamefont {Kuhnert}, \citenamefont {Langen}, \citenamefont {Kitagawa},
  \citenamefont {Rauer}, \citenamefont {Schreitl}, \citenamefont {Mazets},
  \citenamefont {Smith}, \citenamefont {Demler},\ and\ \citenamefont
  {Schmiedmayer}}]{Gring2012}%
  \BibitemOpen
  \bibfield  {author} {\bibinfo {author} {\bibfnamefont {M.}~\bibnamefont
  {Gring}}, \bibinfo {author} {\bibfnamefont {M.}~\bibnamefont {Kuhnert}},
  \bibinfo {author} {\bibfnamefont {T.}~\bibnamefont {Langen}}, \bibinfo
  {author} {\bibfnamefont {T.}~\bibnamefont {Kitagawa}}, \bibinfo {author}
  {\bibfnamefont {B.}~\bibnamefont {Rauer}}, \bibinfo {author} {\bibfnamefont
  {M.}~\bibnamefont {Schreitl}}, \bibinfo {author} {\bibfnamefont
  {I.}~\bibnamefont {Mazets}}, \bibinfo {author} {\bibfnamefont {D.~A.}\
  \bibnamefont {Smith}}, \bibinfo {author} {\bibfnamefont {E.}~\bibnamefont
  {Demler}},\ and\ \bibinfo {author} {\bibfnamefont {J.}~\bibnamefont
  {Schmiedmayer}},\ }\bibfield  {title} {\bibinfo {title} {Relaxation and
  prethermalization in an isolated quantum system},\ }\href {\doibase 10.1126/science.1224953} {\bibfield
  {journal} {\bibinfo  {journal} {Science}\ }\textbf {\bibinfo {volume}
  {337}},\ \bibinfo {pages} {1318} (\bibinfo {year} {2012})}\BibitemShut
  {NoStop}%
\bibitem [{\citenamefont {Gramsch}\ and\ \citenamefont
  {Rigol}(2012)}]{Gramsch2012}%
  \BibitemOpen
  \bibfield  {author} {\bibinfo {author} {\bibfnamefont {C.}~\bibnamefont
  {Gramsch}}\ and\ \bibinfo {author} {\bibfnamefont {M.}~\bibnamefont
  {Rigol}},\ }\bibfield  {title} {\bibinfo {title} {Quenches in a
  quasidisordered integrable lattice system: Dynamics and statistical
  description of observables after relaxation},\ }\href
  {\doibase 10.1103/PhysRevA.86.053615} {\bibfield  {journal} {\bibinfo
  {journal} {Phys. Rev. A}\ }\textbf {\bibinfo {volume} {86}},\ \bibinfo
  {pages} {053615} (\bibinfo {year} {2012})}\BibitemShut {NoStop}%
\bibitem [{\citenamefont {He}\ \emph {et~al.}(2013)\citenamefont {He},
  \citenamefont {Santos}, \citenamefont {Wright},\ and\ \citenamefont
  {Rigol}}]{HeSantos2013}%
  \BibitemOpen
  \bibfield  {author} {\bibinfo {author} {\bibfnamefont {K.}~\bibnamefont
  {He}}, \bibinfo {author} {\bibfnamefont {L.~F.}\ \bibnamefont {Santos}},
  \bibinfo {author} {\bibfnamefont {T.~M.}\ \bibnamefont {Wright}},\ and\
  \bibinfo {author} {\bibfnamefont {M.}~\bibnamefont {Rigol}},\ }\bibfield
  {title} {\bibinfo {title} {Single-particle and many-body analyses of a
  quasiperiodic integrable system after a quench},\ }\href
  {\doibase 10.1103/PhysRevA.87.063637} {\bibfield  {journal} {\bibinfo
  {journal} {Phys. Rev. A}\ }\textbf {\bibinfo {volume} {87}},\ \bibinfo
  {pages} {063637} (\bibinfo {year} {2013})}\BibitemShut {NoStop}%
\bibitem [{\citenamefont {Gong}\ \emph {et~al.}(2021)\citenamefont {Gong},
  \citenamefont {de~Moraes~Neto}, \citenamefont {Zha}, \citenamefont {Wu},
  \citenamefont {Rong}, \citenamefont {Ye}, \citenamefont {Li}, \citenamefont
  {Zhu}, \citenamefont {Wang}, \citenamefont {Zhao} \emph {et~al.}}]{gong2021}%
  \BibitemOpen
  \bibfield  {author} {\bibinfo {author} {\bibfnamefont {M.}~\bibnamefont
  {Gong}}, \bibinfo {author} {\bibfnamefont {G.~D.}\ \bibnamefont
  {de~Moraes~Neto}}, \bibinfo {author} {\bibfnamefont {C.}~\bibnamefont {Zha}},
  \bibinfo {author} {\bibfnamefont {Y.}~\bibnamefont {Wu}}, \bibinfo {author}
  {\bibfnamefont {H.}~\bibnamefont {Rong}}, \bibinfo {author} {\bibfnamefont
  {Y.}~\bibnamefont {Ye}}, \bibinfo {author} {\bibfnamefont {S.}~\bibnamefont
  {Li}}, \bibinfo {author} {\bibfnamefont {Q.}~\bibnamefont {Zhu}}, \bibinfo
  {author} {\bibfnamefont {S.}~\bibnamefont {Wang}}, \bibinfo {author}
  {\bibfnamefont {Y.}~\bibnamefont {Zhao}}, \emph {et~al.},\ }\bibfield
  {title} {\bibinfo {title} {Experimental characterization of the quantum
  many-body localization transition},\ }\href {\doibase 10.1103/PhysRevResearch.3.033043} {\bibfield  {journal} {\bibinfo  {journal} {Phys. Rev. Res.}\ }\textbf {\bibinfo {volume} {3}},\
  \bibinfo {pages} {033043} (\bibinfo {year} {2021})}\BibitemShut {NoStop}%
\bibitem [{\citenamefont {Doggen}\ \emph {et~al.}(2018)\citenamefont {Doggen},
  \citenamefont {Schindler}, \citenamefont {Tikhonov}, \citenamefont {Mirlin},
  \citenamefont {Neupert}, \citenamefont {Polyakov},\ and\ \citenamefont
  {Gornyi}}]{Doggen2018}%
  \BibitemOpen
  \bibfield  {author} {\bibinfo {author} {\bibfnamefont {E.~V.~H.}\
  \bibnamefont {Doggen}}, \bibinfo {author} {\bibfnamefont {F.}~\bibnamefont
  {Schindler}}, \bibinfo {author} {\bibfnamefont {K.~S.}\ \bibnamefont
  {Tikhonov}}, \bibinfo {author} {\bibfnamefont {A.~D.}\ \bibnamefont
  {Mirlin}}, \bibinfo {author} {\bibfnamefont {T.}~\bibnamefont {Neupert}},
  \bibinfo {author} {\bibfnamefont {D.~G.}\ \bibnamefont {Polyakov}},\ and\
  \bibinfo {author} {\bibfnamefont {I.~V.}\ \bibnamefont {Gornyi}},\ }\bibfield
   {title} {\bibinfo {title} {Many-body localization and delocalization in
  large quantum chains},\ }\href {\doibase 10.1103/PhysRevB.98.174202}
  {\bibfield  {journal} {\bibinfo  {journal} {Phys. Rev. B}\ }\textbf {\bibinfo
  {volume} {98}},\ \bibinfo {pages} {174202} (\bibinfo {year}
  {2018})}\BibitemShut {NoStop}%
\bibitem [{\citenamefont {Maksymov}\ \emph {et~al.}(2020)\citenamefont
  {Maksymov}, \citenamefont {Sierant},\ and\ \citenamefont
  {Zakrzewski}}]{maksymov2020}%
  \BibitemOpen
  \bibfield  {author} {\bibinfo {author} {\bibfnamefont {A.}~\bibnamefont
  {Maksymov}}, \bibinfo {author} {\bibfnamefont {P.}~\bibnamefont {Sierant}},\
  and\ \bibinfo {author} {\bibfnamefont {J.}~\bibnamefont {Zakrzewski}},\
  }\bibfield  {title} {\bibinfo {title} {Many-body localization in a
  one-dimensional optical lattice with speckle disorder},\ }\href {\doibase 10.1103/PhysRevB.102.134205}
  {\bibfield  {journal} {\bibinfo  {journal} {Phys. Rev. B}\ }\textbf {\bibinfo
  {volume} {102}},\ \bibinfo {pages} {134205} (\bibinfo {year}
  {2020})}\BibitemShut {NoStop}%
\bibitem [{\citenamefont {Luitz}(2016)}]{luitz2016long}%
  \BibitemOpen
  \bibfield  {author} {\bibinfo {author} {\bibfnamefont {D.~J.}\ \bibnamefont
  {Luitz}},\ }\bibfield  {title} {\bibinfo {title} {Long tail distributions
  near the many-body localization transition},\ }\href {\doibase 10.1103/PhysRevB.93.134201} {\bibfield
  {journal} {\bibinfo  {journal} {Phys. Rev. B}\ }\textbf {\bibinfo {volume}
  {93}},\ \bibinfo {pages} {134201} (\bibinfo {year} {2016})}\BibitemShut
  {NoStop}%
\bibitem [{\citenamefont {Yu}\ \emph {et~al.}(2016)\citenamefont {Yu},
  \citenamefont {Luitz},\ and\ \citenamefont {Clark}}]{Yu2016}%
  \BibitemOpen
  \bibfield  {author} {\bibinfo {author} {\bibfnamefont {X.}~\bibnamefont
  {Yu}}, \bibinfo {author} {\bibfnamefont {D.~J.}\ \bibnamefont {Luitz}},\ and\
  \bibinfo {author} {\bibfnamefont {B.~K.}\ \bibnamefont {Clark}},\ }\bibfield
  {title} {\bibinfo {title} {Bimodal entanglement entropy distribution in the
  many-body localization transition},\ }\href {\doibase 10.1103/PhysRevB.94.184202} {\bibfield  {journal}
  {\bibinfo  {journal} {Phys. Rev. B}\ }\textbf {\bibinfo {volume} {94}},\
  \bibinfo {pages} {184202} (\bibinfo {year} {2016})}\BibitemShut {NoStop}%
\bibitem [{\citenamefont {L.~Mandelstam}(1945)}]{Mandelstam1945}%
  \BibitemOpen
  \bibfield  {author} {\bibinfo {author} {\bibfnamefont {I.~T.}\ \bibnamefont
  {L.~Mandelstam}},\ }\bibfield  {title} {\bibinfo {title} {The uncertainty
  relation between energy and time in nonrelativistic quantum mechanics},\
  }\href {\doibase 10.1007/978-3-642-74626-0_8} {\bibfield  {journal} {\bibinfo  {journal} {J. Phys. USSR}\
  }\textbf {\bibinfo {volume} {9}},\ \bibinfo {pages} {249–254} (\bibinfo
  {year} {1945})}\BibitemShut {NoStop}%
\bibitem [{\citenamefont {Bhattacharyya}(1983)}]{Bhattacharyya1983}%
  \BibitemOpen
  \bibfield  {author} {\bibinfo {author} {\bibfnamefont {K.}~\bibnamefont
  {Bhattacharyya}},\ }\bibfield  {title} {\bibinfo {title} {{Q}uantum decay and
  the {M}andelstam-{T}amm-energy inequality},\ }\href {\doibase 10.1088/0305-4470/16/13/021} {\bibfield
  {journal} {\bibinfo  {journal} {J. Phys. A}\ }\textbf {\bibinfo {volume}
  {16}},\ \bibinfo {pages} {2993} (\bibinfo {year} {1983})}\BibitemShut
  {NoStop}%
\bibitem [{\citenamefont {Khalfin}(1958)}]{Khalfin1958}%
  \BibitemOpen
  \bibfield  {author} {\bibinfo {author} {\bibfnamefont {L.~A.}\ \bibnamefont
  {Khalfin}},\ }\bibfield  {title} {\bibinfo {title} {Contribution to the decay
  theory of a quasi-stationary state},\ }\href {https://www.osti.gov/biblio/4318804} {\bibfield  {journal}
  {\bibinfo  {journal} {Sov. Phys. JETP}\ }\textbf {\bibinfo {volume} {6}},\
  \bibinfo {pages} {1053} (\bibinfo {year} {1958})}\BibitemShut {NoStop}%
\bibitem [{\citenamefont {Misra}\ and\ \citenamefont
  {Sudarshan}(1977)}]{Misra1977}%
  \BibitemOpen
  \bibfield  {author} {\bibinfo {author} {\bibfnamefont {B.}~\bibnamefont
  {Misra}}\ and\ \bibinfo {author} {\bibfnamefont {E.~G.}\ \bibnamefont
  {Sudarshan}},\ }\bibfield  {title} {\bibinfo {title} {The {Z}eno’s paradox
  in quantum theory},\ }\href {\doibase 10.1063/1.523304} {\bibfield  {journal} {\bibinfo
  {journal} {J. Math. Phys.}\ }\textbf {\bibinfo {volume} {18}},\ \bibinfo
  {pages} {756} (\bibinfo {year} {1977})}\BibitemShut {NoStop}%
\bibitem [{\citenamefont {Pfeifer}(1993)}]{Pfeifer1993}%
  \BibitemOpen
  \bibfield  {author} {\bibinfo {author} {\bibfnamefont {P.}~\bibnamefont
  {Pfeifer}},\ }\bibfield  {title} {\bibinfo {title} {How fast can a quantum
  state change with time?},\ }\href {\doibase 10.1103/PhysRevLett.70.3365} {\bibfield  {journal} {\bibinfo
  {journal} {Phys. Rev. Lett.}\ }\textbf {\bibinfo {volume} {70}},\ \bibinfo
  {pages} {3365} (\bibinfo {year} {1993})}\BibitemShut {NoStop}%
\bibitem [{\citenamefont {Ufink}(1993)}]{Ufink1993}%
  \BibitemOpen
  \bibfield  {author} {\bibinfo {author} {\bibfnamefont {J.}~\bibnamefont
  {Ufink}},\ }\bibfield  {title} {\bibinfo {title} {The rate of evolution of a
  quantum state},\ }\href {\doibase 10.1119/1.17368} {\bibfield
  {journal} {\bibinfo  {journal} {Am. J. Phys.}\ }\textbf {\bibinfo {volume}
  {61}},\ \bibinfo {pages} {935} (\bibinfo {year} {1993})}\BibitemShut
  {NoStop}%
\bibitem [{\citenamefont {Margolus}\ and\ \citenamefont
  {Levitin}(1998)}]{Margolus1998}%
  \BibitemOpen
  \bibfield  {author} {\bibinfo {author} {\bibfnamefont {N.}~\bibnamefont
  {Margolus}}\ and\ \bibinfo {author} {\bibfnamefont {L.~B.}\ \bibnamefont
  {Levitin}},\ }\bibfield  {title} {\bibinfo {title} {The maximum speed of
  dynamical evolution},\ }\href {\doibase 10.1016/S0167-2789(98)00054-2} {\bibfield  {journal} {\bibinfo
  {journal} {Phys. D: Nonlinear Phenom.}\ }\textbf {\bibinfo {volume} {120}},\
  \bibinfo {pages} {188} (\bibinfo {year} {1998})}\BibitemShut {NoStop}%
\bibitem [{\citenamefont {Giovannetti}\ \emph {et~al.}(2003)\citenamefont
  {Giovannetti}, \citenamefont {Lloyd},\ and\ \citenamefont
  {Maccone}}]{Giovannetti2003}%
  \BibitemOpen
  \bibfield  {author} {\bibinfo {author} {\bibfnamefont {V.}~\bibnamefont
  {Giovannetti}}, \bibinfo {author} {\bibfnamefont {S.}~\bibnamefont {Lloyd}},\
  and\ \bibinfo {author} {\bibfnamefont {L.}~\bibnamefont {Maccone}},\
  }\bibfield  {title} {\bibinfo {title} {Quantum limits to dynamical
  evolution},\ }\href {\doibase 10.1103/PhysRevA.67.052109} {\bibfield  {journal} {\bibinfo  {journal} {Phys.
  Rev. A}\ }\textbf {\bibinfo {volume} {67}},\ \bibinfo {pages} {052109}
  (\bibinfo {year} {2003})}\BibitemShut {NoStop}%
\bibitem [{\citenamefont {Schofield}\ \emph {et~al.}(1995)\citenamefont
  {Schofield}, \citenamefont {Wolynes},\ and\ \citenamefont
  {Wyatt}}]{Schofield1995}%
  \BibitemOpen
  \bibfield  {author} {\bibinfo {author} {\bibfnamefont {S.~A.}\ \bibnamefont
  {Schofield}}, \bibinfo {author} {\bibfnamefont {P.~G.}\ \bibnamefont
  {Wolynes}},\ and\ \bibinfo {author} {\bibfnamefont {R.~E.}\ \bibnamefont
  {Wyatt}},\ }\bibfield  {title} {\bibinfo {title} {Computational study of
  many-dimensional quantum energy flow: From action diffusion to
  localization},\ }\href {\doibase 10.1103/PhysRevLett.74.3720} {\bibfield  {journal} {\bibinfo  {journal}
  {Phys. Rev. Lett.}\ }\textbf {\bibinfo {volume} {74}},\ \bibinfo {pages}
  {3720} (\bibinfo {year} {1995})}\BibitemShut {NoStop}%
\bibitem [{\citenamefont {De~Oliveira}\ and\ \citenamefont
  {Pellegrino}(1999)}]{de1999quantum}%
  \BibitemOpen
  \bibfield  {author} {\bibinfo {author} {\bibfnamefont {C.~R.}\ \bibnamefont
  {De~Oliveira}}\ and\ \bibinfo {author} {\bibfnamefont {G.~Q.}\ \bibnamefont
  {Pellegrino}},\ }\bibfield  {title} {\bibinfo {title} {Quantum return
  probability for substitution potentials},\ }\href {\doibase 10.1088/0305-4470/32/26/102} {\bibfield
  {journal} {\bibinfo  {journal} {J. Phys. A Math. Gen.}\ }\textbf {\bibinfo
  {volume} {32}},\ \bibinfo {pages} {L285} (\bibinfo {year}
  {1999})}\BibitemShut {NoStop}%
\bibitem [{\citenamefont {Miyamoto}(2001)}]{Miyamoto2001}%
  \BibitemOpen
  \bibfield  {author} {\bibinfo {author} {\bibfnamefont {M.}~\bibnamefont
  {Miyamoto}},\ }\bibfield  {title} {\bibinfo {title} {A generalized {W}eyl
  relation approach to the time operator and its connection to the survival
  probability},\ }\href {\doibase 10.1063/1.1346598} {\bibfield  {journal} {\bibinfo  {journal} {J.
  Math. Phys.}\ }\textbf {\bibinfo {volume} {42}},\ \bibinfo {pages} {1038}
  (\bibinfo {year} {2001})}\BibitemShut {NoStop}%
\bibitem [{\citenamefont {Krauss}\ and\ \citenamefont
  {Dent}(2008)}]{krauss2008late}%
  \BibitemOpen
  \bibfield  {author} {\bibinfo {author} {\bibfnamefont {L.~M.}\ \bibnamefont
  {Krauss}}\ and\ \bibinfo {author} {\bibfnamefont {J.}~\bibnamefont {Dent}},\
  }\bibfield  {title} {\bibinfo {title} {Late time behavior of false vacuum
  decay: Possible implications for cosmology and metastable inflating states},\
  }\href {\doibase 10.1103/PhysRevLett.100.171301}{\bibfield {journal}{\bibinfo {journal}{Phys. Rev. Lett.}\
  }\textbf {\bibinfo {volume} {100}},\ \bibinfo {pages} {171301} (\bibinfo
  {year} {2008})}\BibitemShut {NoStop}%
\bibitem [{\citenamefont {Xu}\ and\ \citenamefont {Liu}(2008)}]{xu2008}%
  \BibitemOpen
  \bibfield  {author} {\bibinfo {author} {\bibfnamefont {X.}~\bibnamefont
  {Xu}}\ and\ \bibinfo {author} {\bibfnamefont {F.}~\bibnamefont {Liu}},\
  }\bibfield  {title} {\bibinfo {title} {Continuous-time quantum walks on
  {E}rd{\"o}s--{R}{\'e}nyi networks},\ }\href {\doibase 10.1016/j.physleta.2008.09.042} {\bibfield  {journal}
  {\bibinfo  {journal} {Phys. Lett. A}\ }\textbf {\bibinfo {volume} {372}},\
  \bibinfo {pages} {6727} (\bibinfo {year} {2008})}\BibitemShut {NoStop}%
\bibitem [{\citenamefont {M{\"u}lken}\ and\ \citenamefont
  {Blumen}(2011)}]{mulken2011}%
  \BibitemOpen
  \bibfield  {author} {\bibinfo {author} {\bibfnamefont {O.}~\bibnamefont
  {M{\"u}lken}}\ and\ \bibinfo {author} {\bibfnamefont {A.}~\bibnamefont
  {Blumen}},\ }\bibfield  {title} {\bibinfo {title} {Continuous-time quantum
  walks: Models for coherent transport on complex networks},\ }\href {\doibase 10.1016/j.physrep.2011.01.002}
  {\bibfield  {journal} {\bibinfo  {journal} {Phys. Rep.}\ }\textbf {\bibinfo
  {volume} {502}},\ \bibinfo {pages} {37} (\bibinfo {year} {2011})}\BibitemShut
  {NoStop}%
\bibitem [{\citenamefont {Ampadu}(2012)}]{ampadu2012return}%
  \BibitemOpen
  \bibfield  {author} {\bibinfo {author} {\bibfnamefont {C.}~\bibnamefont
  {Ampadu}},\ }\bibfield  {title} {\bibinfo {title} {Return probability of the
  {F}ibonacci quantum walk},\ }\href {\doibase 10.1088/0253-6102/58/2/09} {\bibfield  {journal} {\bibinfo
  {journal} {Commun. Theor. Phys.}\ }\textbf {\bibinfo {volume} {58}},\
  \bibinfo {pages} {220} (\bibinfo {year} {2012})}\BibitemShut {NoStop}%
\bibitem [{\citenamefont {Riascos}\ and\ \citenamefont
  {Mateos}(2015)}]{riascos2015}%
  \BibitemOpen
  \bibfield  {author} {\bibinfo {author} {\bibfnamefont {A.}~\bibnamefont
  {Riascos}}\ and\ \bibinfo {author} {\bibfnamefont {J.~L.}\ \bibnamefont
  {Mateos}},\ }\bibfield  {title} {\bibinfo {title} {Fractional quantum
  mechanics on networks: Long-range dynamics and quantum transport},\
  }\href {\doibase 10.1103/PhysRevE.92.052814} {\bibfield  {journal} {\bibinfo  {journal} {Phys. Rev. E}\
  }\textbf {\bibinfo {volume} {92}},\ \bibinfo {pages} {052814} (\bibinfo
  {year} {2015})}\BibitemShut {NoStop}%
\bibitem [{\citenamefont {del Campo}\ \emph {et~al.}(2017)\citenamefont {del
  Campo}, \citenamefont {Molina-Vilaplana},\ and\ \citenamefont
  {Sonner}}]{Campo2017}%
  \BibitemOpen
  \bibfield  {author} {\bibinfo {author} {\bibfnamefont {A.}~\bibnamefont {del
  Campo}}, \bibinfo {author} {\bibfnamefont {J.}~\bibnamefont
  {Molina-Vilaplana}},\ and\ \bibinfo {author} {\bibfnamefont {J.}~\bibnamefont
  {Sonner}},\ }\bibfield  {title} {\bibinfo {title} {Scrambling the spectral
  form factor: Unitarity constraints and exact results},\ }\href
  {\doibase 10.1103/PhysRevD.95.126008} {\bibfield  {journal} {\bibinfo
  {journal} {Phys. Rev. D}\ }\textbf {\bibinfo {volume} {95}},\ \bibinfo
  {pages} {126008} (\bibinfo {year} {2017})}\BibitemShut {NoStop}%
\bibitem [{\citenamefont {del Campo}\ \emph {et~al.}(2018)\citenamefont {del
  Campo}, \citenamefont {Molina-Vilaplana}, \citenamefont {Santos},\ and\
  \citenamefont {Sonner}}]{CampoSantos2018}%
  \BibitemOpen
  \bibfield  {author} {\bibinfo {author} {\bibfnamefont {A.}~\bibnamefont {del
  Campo}}, \bibinfo {author} {\bibfnamefont {J.}~\bibnamefont
  {Molina-Vilaplana}}, \bibinfo {author} {\bibfnamefont {L.~F.}\ \bibnamefont
  {Santos}},\ and\ \bibinfo {author} {\bibfnamefont {J.}~\bibnamefont
  {Sonner}},\ }\bibfield  {title} {\bibinfo {title} {Decay of a
  thermofield-double state in chaotic quantum systems},\ }\href
  {\doibase 10.1140/epjst/e2018-00083-5} {\bibfield  {journal} {\bibinfo
   {journal} {Eur. Phys. J. Spec. Top.}\ }\textbf {\bibinfo {volume} {227}},\
  \bibinfo {pages} {247} (\bibinfo {year} {2018})}\BibitemShut {NoStop}%
\bibitem [{\citenamefont {Lerma-Hern{\'a}ndez}\ \emph
  {et~al.}(2019)\citenamefont {Lerma-Hern{\'a}ndez}, \citenamefont
  {Villase{\~n}or}, \citenamefont {Bastarrachea-Magnani}, \citenamefont
  {Torres-Herrera}, \citenamefont {Santos},\ and\ \citenamefont
  {Hirsch}}]{lerma2019}%
  \BibitemOpen
  \bibfield  {author} {\bibinfo {author} {\bibfnamefont {S.}~\bibnamefont
  {Lerma-Hern{\'a}ndez}}, \bibinfo {author} {\bibfnamefont {D.}~\bibnamefont
  {Villase{\~n}or}}, \bibinfo {author} {\bibfnamefont {M.}~\bibnamefont
  {Bastarrachea-Magnani}}, \bibinfo {author} {\bibfnamefont {E.}~\bibnamefont
  {Torres-Herrera}}, \bibinfo {author} {\bibfnamefont {L.~F.}\ \bibnamefont
  {Santos}},\ and\ \bibinfo {author} {\bibfnamefont {J.}~\bibnamefont
  {Hirsch}},\ }\bibfield  {title} {\bibinfo {title} {Dynamical signatures of
  quantum chaos and relaxation timescales in a spin-boson system},\
  }\href {\doibase 10.1103/PhysRevE.100.012218} {\bibfield  {journal} {\bibinfo  {journal} {Phys. Rev. E}\
  }\textbf {\bibinfo {volume} {100}},\ \bibinfo {pages} {012218} (\bibinfo
  {year} {2019})}\BibitemShut {NoStop}%
\bibitem [{\citenamefont {de~la Cruz}\ \emph {et~al.}(2020)\citenamefont {de~la
  Cruz}, \citenamefont {Lerma-Hern{\'a}ndez},\ and\ \citenamefont
  {Hirsch}}]{delacruz2020}%
  \BibitemOpen
  \bibfield  {author} {\bibinfo {author} {\bibfnamefont {J.}~\bibnamefont
  {de~la Cruz}}, \bibinfo {author} {\bibfnamefont {S.}~\bibnamefont
  {Lerma-Hern{\'a}ndez}},\ and\ \bibinfo {author} {\bibfnamefont {J.~G.}\
  \bibnamefont {Hirsch}},\ }\bibfield  {title} {\bibinfo {title} {Quantum chaos
  in a system with high degree of symmetries},\ }\href {\doibase 10.1103/PhysRevE.102.032208} {\bibfield
  {journal} {\bibinfo  {journal} {Phys. Rev. E}\ }\textbf {\bibinfo {volume}
  {102}},\ \bibinfo {pages} {032208} (\bibinfo {year} {2020})}\BibitemShut
  {NoStop}%
\bibitem [{\citenamefont {Leviandier}\ \emph {et~al.}(1986)\citenamefont
  {Leviandier}, \citenamefont {Lombardi}, \citenamefont {Jost},\ and\
  \citenamefont {Pique}}]{Leviandier1986}%
  \BibitemOpen
  \bibfield  {author} {\bibinfo {author} {\bibfnamefont {L.}~\bibnamefont
  {Leviandier}}, \bibinfo {author} {\bibfnamefont {M.}~\bibnamefont
  {Lombardi}}, \bibinfo {author} {\bibfnamefont {R.}~\bibnamefont {Jost}},\
  and\ \bibinfo {author} {\bibfnamefont {J.~P.}\ \bibnamefont {Pique}},\
  }\bibfield  {title} {\bibinfo {title} {Fourier transform: A tool to measure
  statistical level properties in very complex spectra},\ }\href
  {\doibase 10.1103/PhysRevLett.56.2449} {\bibfield  {journal} {\bibinfo
   {journal} {Phys. Rev. Lett.}\ }\textbf {\bibinfo {volume} {56}},\ \bibinfo
  {pages} {2449} (\bibinfo {year} {1986})}\BibitemShut {NoStop}%
\bibitem [{\citenamefont {Gruebele}\ and\ \citenamefont
  {Wolynes}(2004)}]{Gruebele2004}%
  \BibitemOpen
  \bibfield  {author} {\bibinfo {author} {\bibfnamefont {M.}~\bibnamefont
  {Gruebele}}\ and\ \bibinfo {author} {\bibfnamefont {P.~G.}\ \bibnamefont
  {Wolynes}},\ }\bibfield  {title} {\bibinfo {title} {Vibrational energy flow
  and chemical reactions},\ }\href {\doibase 10.1021/ar030230t} {\bibfield  {journal} {\bibinfo
  {journal} {Acc. Chem. Res.}\ }\textbf {\bibinfo {volume} {37}},\ \bibinfo
  {pages} {261} (\bibinfo {year} {2004})}\BibitemShut {NoStop}%
\bibitem [{\citenamefont {Rothe}\ \emph {et~al.}(2006)\citenamefont {Rothe},
  \citenamefont {Hintschich},\ and\ \citenamefont
  {Monkman}}]{rothe2006violation}%
  \BibitemOpen
  \bibfield  {author} {\bibinfo {author} {\bibfnamefont {C.}~\bibnamefont
  {Rothe}}, \bibinfo {author} {\bibfnamefont {S.}~\bibnamefont {Hintschich}},\
  and\ \bibinfo {author} {\bibfnamefont {A.}~\bibnamefont {Monkman}},\
  }\bibfield  {title} {\bibinfo {title} {Violation of the exponential-decay law
  at long times},\ }\href {\doibase 10.1103/PhysRevLett.96.163601} {\bibfield  {journal} {\bibinfo  {journal}
  {Phys. Rev. Lett.}\ }\textbf {\bibinfo {volume} {96}},\ \bibinfo {pages}
  {163601} (\bibinfo {year} {2006})}\BibitemShut {NoStop}%
\bibitem [{\citenamefont {Wilkinson}\ \emph {et~al.}(1997)\citenamefont
  {Wilkinson}, \citenamefont {Bharucha}, \citenamefont {Fischer}, \citenamefont
  {Madison}, \citenamefont {Morrow}, \citenamefont {Niu}, \citenamefont
  {Sundaram},\ and\ \citenamefont {Raizen}}]{wilkinson1997}%
  \BibitemOpen
  \bibfield  {author} {\bibinfo {author} {\bibfnamefont {S.~R.}\ \bibnamefont
  {Wilkinson}}, \bibinfo {author} {\bibfnamefont {C.~F.}\ \bibnamefont
  {Bharucha}}, \bibinfo {author} {\bibfnamefont {M.~C.}\ \bibnamefont
  {Fischer}}, \bibinfo {author} {\bibfnamefont {K.~W.}\ \bibnamefont
  {Madison}}, \bibinfo {author} {\bibfnamefont {P.~R.}\ \bibnamefont {Morrow}},
  \bibinfo {author} {\bibfnamefont {Q.}~\bibnamefont {Niu}}, \bibinfo {author}
  {\bibfnamefont {B.}~\bibnamefont {Sundaram}},\ and\ \bibinfo {author}
  {\bibfnamefont {M.~G.}\ \bibnamefont {Raizen}},\ }\bibfield  {title}
  {\bibinfo {title} {Experimental evidence for non-exponential decay in quantum
  tunnelling},\ }\href {\doibase 10.1038/42418} {\bibfield  {journal} {\bibinfo  {journal}
  {Nature}\ }\textbf {\bibinfo {volume} {387}},\ \bibinfo {pages} {575}
  (\bibinfo {year} {1997})}\BibitemShut {NoStop}%
\bibitem [{\citenamefont {Gherardini}\ \emph {et~al.}(2017)\citenamefont
  {Gherardini}, \citenamefont {Lovecchio}, \citenamefont {M{\"u}ller},
  \citenamefont {Lombardi}, \citenamefont {Caruso},\ and\ \citenamefont
  {Cataliotti}}]{gherardini2017}%
  \BibitemOpen
  \bibfield  {author} {\bibinfo {author} {\bibfnamefont {S.}~\bibnamefont
  {Gherardini}}, \bibinfo {author} {\bibfnamefont {C.}~\bibnamefont
  {Lovecchio}}, \bibinfo {author} {\bibfnamefont {M.~M.}\ \bibnamefont
  {M{\"u}ller}}, \bibinfo {author} {\bibfnamefont {P.}~\bibnamefont
  {Lombardi}}, \bibinfo {author} {\bibfnamefont {F.}~\bibnamefont {Caruso}},\
  and\ \bibinfo {author} {\bibfnamefont {F.~S.}\ \bibnamefont {Cataliotti}},\
  }\bibfield  {title} {\bibinfo {title} {Ergodicity in randomly perturbed
  quantum systems},\ }\href {\doibase 10.1088/2058-9565/aa5d00} {\bibfield  {journal} {\bibinfo  {journal}
  {Quantum Sci. Technol.}\ }\textbf {\bibinfo {volume} {2}},\ \bibinfo {pages}
  {015007} (\bibinfo {year} {2017})}\BibitemShut {NoStop}%
\bibitem [{\citenamefont {Singh}\ \emph {et~al.}(2019)\citenamefont {Singh},
  \citenamefont {Fujiwara}, \citenamefont {Geiger}, \citenamefont {Simmons},
  \citenamefont {Lipatov}, \citenamefont {Cao}, \citenamefont {Dotti},
  \citenamefont {Rajagopal}, \citenamefont {Senaratne}, \citenamefont
  {Shimasaki} \emph {et~al.}}]{singh2019}%
  \BibitemOpen
  \bibfield  {author} {\bibinfo {author} {\bibfnamefont {K.}~\bibnamefont
  {Singh}}, \bibinfo {author} {\bibfnamefont {C.~J.}\ \bibnamefont {Fujiwara}},
  \bibinfo {author} {\bibfnamefont {Z.~A.}\ \bibnamefont {Geiger}}, \bibinfo
  {author} {\bibfnamefont {E.~Q.}\ \bibnamefont {Simmons}}, \bibinfo {author}
  {\bibfnamefont {M.}~\bibnamefont {Lipatov}}, \bibinfo {author} {\bibfnamefont
  {A.}~\bibnamefont {Cao}}, \bibinfo {author} {\bibfnamefont {P.}~\bibnamefont
  {Dotti}}, \bibinfo {author} {\bibfnamefont {S.~V.}\ \bibnamefont
  {Rajagopal}}, \bibinfo {author} {\bibfnamefont {R.}~\bibnamefont
  {Senaratne}}, \bibinfo {author} {\bibfnamefont {T.}~\bibnamefont
  {Shimasaki}}, \emph {et~al.},\ }\bibfield  {title} {\bibinfo {title}
  {Quantifying and controlling prethermal nonergodicity in interacting
  {F}loquet matter},\ }\href {\doibase 10.1103/PhysRevX.9.041021} {\bibfield  {journal} {\bibinfo  {journal}
  {Phys. Rev. X}\ }\textbf {\bibinfo {volume} {9}},\ \bibinfo {pages} {041021}
  (\bibinfo {year} {2019})}\BibitemShut {NoStop}%
\bibitem [{\citenamefont {Jordan}\ and\ \citenamefont
  {Wigner}(1928)}]{Jordan_1928}%
  \BibitemOpen
  \bibfield  {author} {\bibinfo {author} {\bibfnamefont {P.}~\bibnamefont
  {Jordan}}\ and\ \bibinfo {author} {\bibfnamefont {E.~P.}\ \bibnamefont
  {Wigner}},\ }\bibfield  {title} {\bibinfo {title} {{\"U}ber das {P}aulische
  {\"a}quivalenzverbot},\ }\href {\doibase 10.1007/BF01331938}
  {\bibfield  {journal} {\bibinfo  {journal} {Z. Phys.}\ }\textbf {\bibinfo
  {volume} {47}},\ \bibinfo {pages} {631} (\bibinfo {year} {1928})}\BibitemShut
  {NoStop}%
\bibitem [{\citenamefont {Lieb}\ \emph {et~al.}(1961)\citenamefont {Lieb},
  \citenamefont {Schultz},\ and\ \citenamefont {Mattis}}]{Lieb1961}%
  \BibitemOpen
  \bibfield  {author} {\bibinfo {author} {\bibfnamefont {E.}~\bibnamefont
  {Lieb}}, \bibinfo {author} {\bibfnamefont {T.}~\bibnamefont {Schultz}},\ and\
  \bibinfo {author} {\bibfnamefont {D.}~\bibnamefont {Mattis}},\ }\bibfield
  {title} {\bibinfo {title} {Two soluble models of an antiferromagnetic
  chain},\ }\href {\doibase 10.1016/0003-4916(61)90115-4} {\bibfield  {journal} {\bibinfo  {journal} {Ann.
  Phys.}\ }\textbf {\bibinfo {volume} {16}},\ \bibinfo {pages} {407} (\bibinfo
  {year} {1961})}\BibitemShut {NoStop}%
\bibitem [{\citenamefont {Bethe}(1931)}]{Bethe1931}%
  \BibitemOpen
  \bibfield  {author} {\bibinfo {author} {\bibfnamefont {H.}~\bibnamefont
  {Bethe}},\ }\bibfield  {title} {\bibinfo {title} {Zur theorie der metalle},\
  }\href {\doibase 10.1007/BF01341708} {\bibfield  {journal} {\bibinfo  {journal} {Z. Phys.}\
  }\textbf {\bibinfo {volume} {71}},\ \bibinfo {pages} {205} (\bibinfo {year}
  {1931})}\BibitemShut {NoStop}%
\bibitem [{\citenamefont {Hulth{\'e}n}(1938)}]{hulthen1938}%
  \BibitemOpen
  \bibfield  {author} {\bibinfo {author} {\bibfnamefont {L.}~\bibnamefont
  {Hulth{\'e}n}},\ }\emph {\bibinfo {title} {{\"U}ber das austauschproblem
  eines kristalles}},\ \href {\doibase } {Ph.D. thesis},\ \bibinfo  {school}
  {Almqvist \& Wiksell} (\bibinfo {year} {1938})\BibitemShut {NoStop}%
\bibitem [{\citenamefont {Des~Cloizeaux}\ and\ \citenamefont
  {Pearson}(1962)}]{des1962}%
  \BibitemOpen
  \bibfield  {author} {\bibinfo {author} {\bibfnamefont {J.}~\bibnamefont
  {Des~Cloizeaux}}\ and\ \bibinfo {author} {\bibfnamefont {J.}~\bibnamefont
  {Pearson}},\ }\bibfield  {title} {\bibinfo {title} {Spin-wave spectrum of the
  antiferromagnetic linear chain},\ }\href {\doibase 10.1103/PhysRev.128.2131} {\bibfield  {journal}
  {\bibinfo  {journal} {Phys. Rev.}\ }\textbf {\bibinfo {volume} {128}},\
  \bibinfo {pages} {2131} (\bibinfo {year} {1962})}\BibitemShut {NoStop}%
\bibitem [{\citenamefont {Yang}\ and\ \citenamefont
  {Yang}(1966)}]{yang1966one}%
  \BibitemOpen
  \bibfield  {author} {\bibinfo {author} {\bibfnamefont {C.-N.}\ \bibnamefont
  {Yang}}\ and\ \bibinfo {author} {\bibfnamefont {C.-P.}\ \bibnamefont
  {Yang}},\ }\bibfield  {title} {\bibinfo {title} {One-dimensional chain of
  anisotropic spin-spin interactions. I. Proof of {B}ethe's hypothesis for
  ground state in a finite system},\ }\href {\doibase 10.1103/PhysRev.150.321} {\bibfield  {journal}
  {\bibinfo  {journal} {Phys. Rev.}\ }\textbf {\bibinfo {volume} {150}},\
  \bibinfo {pages} {321} (\bibinfo {year} {1966})}\BibitemShut {NoStop}%
\bibitem [{\citenamefont {Kirillov}\ and\ \citenamefont
  {Reshetikhin}(1987)}]{kirillov1987I}%
  \BibitemOpen
  \bibfield  {author} {\bibinfo {author} {\bibfnamefont {A.~N.}\ \bibnamefont
  {Kirillov}}\ and\ \bibinfo {author} {\bibfnamefont {N.~Y.}\ \bibnamefont
  {Reshetikhin}},\ }\bibfield  {title} {\bibinfo {title} {Exact solution of the
  integrable {$XXZ$} {H}eisenberg model with arbitrary spin. I. The ground state
  and the excitation spectrum},\ }\href {\doibase 10.1088/0305-4470/20/6/038} {\bibfield  {journal} {\bibinfo
   {journal} {J. Phys. A Math. Gen.}\ }\textbf {\bibinfo {volume} {20}},\
  \bibinfo {pages} {1565} (\bibinfo {year} {1987})}\BibitemShut {NoStop}%
\bibitem [{\citenamefont {Oganesyan}\ and\ \citenamefont
  {Huse}(2007)}]{Oganesyan2007}%
  \BibitemOpen
  \bibfield  {author} {\bibinfo {author} {\bibfnamefont {V.}~\bibnamefont
  {Oganesyan}}\ and\ \bibinfo {author} {\bibfnamefont {D.~A.}\ \bibnamefont
  {Huse}},\ }\bibfield  {title} {\bibinfo {title} {Localization of interacting
  fermions at high temperature},\ }\href {\doibase 10.1103/PhysRevB.75.155111} {\bibfield  {journal}
  {\bibinfo  {journal} {Phys. Rev. B}\ }\textbf {\bibinfo {volume} {75}},\
  \bibinfo {pages} {155111} (\bibinfo {year} {2007})}\BibitemShut {NoStop}%
\bibitem [{\citenamefont {Pal}\ and\ \citenamefont {Huse}(2010)}]{Pal2010}%
  \BibitemOpen
  \bibfield  {author} {\bibinfo {author} {\bibfnamefont {A.}~\bibnamefont
  {Pal}}\ and\ \bibinfo {author} {\bibfnamefont {D.~A.}\ \bibnamefont {Huse}},\
  }\bibfield  {title} {\bibinfo {title} {Many-body localization phase
  transition},\ }\href {\doibase 10.1103/PhysRevB.82.174411} {\bibfield
  {journal} {\bibinfo  {journal} {Phys. Rev. B}\ }\textbf {\bibinfo {volume}
  {82}},\ \bibinfo {pages} {174411} (\bibinfo {year} {2010})}\BibitemShut
  {NoStop}%
\bibitem [{\citenamefont {Berkelbach}\ and\ \citenamefont
  {Reichman}(2010)}]{Berkelbach2010}%
  \BibitemOpen
  \bibfield  {author} {\bibinfo {author} {\bibfnamefont {T.~C.}\ \bibnamefont
  {Berkelbach}}\ and\ \bibinfo {author} {\bibfnamefont {D.~R.}\ \bibnamefont
  {Reichman}},\ }\bibfield  {title} {\bibinfo {title} {Conductivity of
  disordered quantum lattice models at infinite temperature: {M}any-body
  localization},\ }\href {\doibase 10.1103/PhysRevB.81.224429} {\bibfield  {journal} {\bibinfo  {journal}
  {Phys. Rev. B}\ }\textbf {\bibinfo {volume} {81}},\ \bibinfo {pages} {224429}
  (\bibinfo {year} {2010})}\BibitemShut {NoStop}%
\bibitem [{\citenamefont {Kj{\"a}ll}\ \emph {et~al.}(2014)\citenamefont
  {Kj{\"a}ll}, \citenamefont {Bardarson},\ and\ \citenamefont
  {Pollmann}}]{Kjall2014}%
  \BibitemOpen
  \bibfield  {author} {\bibinfo {author} {\bibfnamefont {J.~A.}\ \bibnamefont
  {Kj{\"a}ll}}, \bibinfo {author} {\bibfnamefont {J.~H.}\ \bibnamefont
  {Bardarson}},\ and\ \bibinfo {author} {\bibfnamefont {F.}~\bibnamefont
  {Pollmann}},\ }\bibfield  {title} {\bibinfo {title} {Many-body localization
  in a disordered quantum {I}sing chain},\ }\href {\doibase 10.1103/PhysRevLett.113.107204} {\bibfield  {journal}
  {\bibinfo  {journal} {Phys. Rev. Lett.}\ }\textbf {\bibinfo {volume} {113}},\
  \bibinfo {pages} {107204} (\bibinfo {year} {2014})}\BibitemShut {NoStop}%
\bibitem [{\citenamefont {Luitz}\ \emph {et~al.}(2015)\citenamefont {Luitz},
  \citenamefont {Laflorencie},\ and\ \citenamefont {Alet}}]{Luitz2015}%
  \BibitemOpen
  \bibfield  {author} {\bibinfo {author} {\bibfnamefont {D.~J.}\ \bibnamefont
  {Luitz}}, \bibinfo {author} {\bibfnamefont {N.}~\bibnamefont {Laflorencie}},\
  and\ \bibinfo {author} {\bibfnamefont {F.}~\bibnamefont {Alet}},\ }\bibfield
  {title} {\bibinfo {title} {Many-body localization edge in the random-field
  {H}eisenberg chain},\ }\href {\doibase 10.1103/PhysRevB.91.081103}
  {\bibfield  {journal} {\bibinfo  {journal} {Phys. Rev. B}\ }\textbf {\bibinfo
  {volume} {91}},\ \bibinfo {pages} {081103} (\bibinfo {year}
  {2015})}\BibitemShut {NoStop}%
\bibitem [{\citenamefont {Devakul}\ and\ \citenamefont
  {Singh}(2015)}]{Devakul2015}%
  \BibitemOpen
  \bibfield  {author} {\bibinfo {author} {\bibfnamefont {T.}~\bibnamefont
  {Devakul}}\ and\ \bibinfo {author} {\bibfnamefont {R.~R.}\ \bibnamefont
  {Singh}},\ }\bibfield  {title} {\bibinfo {title} {Early breakdown of area-law
  entanglement at the many-body delocalization transition},\ }\href {\doibase 10.1103/PhysRevLett.115.187201}
  {\bibfield  {journal} {\bibinfo  {journal} {Phys. Rev. Lett.}\ }\textbf
  {\bibinfo {volume} {115}},\ \bibinfo {pages} {187201} (\bibinfo {year}
  {2015})}\BibitemShut {NoStop}%
\bibitem [{\citenamefont {Sierant}\ \emph {et~al.}(2020)\citenamefont
  {Sierant}, \citenamefont {Delande},\ and\ \citenamefont
  {Zakrzewski}}]{Sierant_2020}%
  \BibitemOpen
  \bibfield  {author} {\bibinfo {author} {\bibfnamefont {P.}~\bibnamefont
  {Sierant}}, \bibinfo {author} {\bibfnamefont {D.}~\bibnamefont {Delande}},\
  and\ \bibinfo {author} {\bibfnamefont {J.}~\bibnamefont {Zakrzewski}},\
  }\bibfield  {title} {\bibinfo {title} {Thouless time analysis of {A}nderson
  and many-body localization transitions},\ }\href
  {\doibase 10.1103/PhysRevLett.124.186601} {\bibfield  {journal}
  {\bibinfo  {journal} {Phys. Rev. Lett.}\ }\textbf {\bibinfo {volume} {124}},\
  \bibinfo {pages} {186601} (\bibinfo {year} {2020})}\BibitemShut {NoStop}%
\bibitem [{\citenamefont {Schreiber}\ \emph {et~al.}(2015)\citenamefont
  {Schreiber}, \citenamefont {Hodgman}, \citenamefont {Bordia}, \citenamefont
  {L{\"u}schen}, \citenamefont {Fischer}, \citenamefont {Vosk}, \citenamefont
  {Altman}, \citenamefont {Schneider},\ and\ \citenamefont
  {Bloch}}]{Schreiber2015}%
  \BibitemOpen
  \bibfield  {author} {\bibinfo {author} {\bibfnamefont {M.}~\bibnamefont
  {Schreiber}}, \bibinfo {author} {\bibfnamefont {S.~S.}\ \bibnamefont
  {Hodgman}}, \bibinfo {author} {\bibfnamefont {P.}~\bibnamefont {Bordia}},
  \bibinfo {author} {\bibfnamefont {H.~P.}\ \bibnamefont {L{\"u}schen}},
  \bibinfo {author} {\bibfnamefont {M.~H.}\ \bibnamefont {Fischer}}, \bibinfo
  {author} {\bibfnamefont {R.}~\bibnamefont {Vosk}}, \bibinfo {author}
  {\bibfnamefont {E.}~\bibnamefont {Altman}}, \bibinfo {author} {\bibfnamefont
  {U.}~\bibnamefont {Schneider}},\ and\ \bibinfo {author} {\bibfnamefont
  {I.}~\bibnamefont {Bloch}},\ }\bibfield  {title} {\bibinfo {title}
  {Observation of many-body localization of interacting fermions in a
  quasirandom optical lattice},\ }\href
  {\doibase 10.1126/science.aaa7432} {\bibfield  {journal} {\bibinfo
  {journal} {Science}\ }\textbf {\bibinfo {volume} {349}},\ \bibinfo {pages}
  {842} (\bibinfo {year} {2015})}\BibitemShut {NoStop}%
\bibitem [{\citenamefont {Bordia}\ \emph {et~al.}(2017)\citenamefont {Bordia},
  \citenamefont {L\"{u}schen}, \citenamefont {Schneider}, \citenamefont
  {Knap},\ and\ \citenamefont {Bloch}}]{Bordia2017a}%
  \BibitemOpen
  \bibfield  {author} {\bibinfo {author} {\bibfnamefont {P.}~\bibnamefont
  {Bordia}}, \bibinfo {author} {\bibfnamefont {H.}~\bibnamefont {L\"{u}schen}},
  \bibinfo {author} {\bibfnamefont {U.}~\bibnamefont {Schneider}}, \bibinfo
  {author} {\bibfnamefont {M.}~\bibnamefont {Knap}},\ and\ \bibinfo {author}
  {\bibfnamefont {I.}~\bibnamefont {Bloch}},\ }\bibfield  {title} {\bibinfo
  {title} {Periodically driving a many-body localized quantum system},\ }\href
  {\doibase 10.1038/nphys4020} {\bibfield  {journal} {\bibinfo
  {journal} {Nat. Phys.}\ }\textbf {\bibinfo {volume} {13}},\ \bibinfo {pages}
  {460} (\bibinfo {year} {2017})}\BibitemShut {NoStop}%
\bibitem [{\citenamefont {Richerme}\ \emph {et~al.}(2014)\citenamefont
  {Richerme}, \citenamefont {Gong}, \citenamefont {Lee}, \citenamefont {Senko},
  \citenamefont {Smith}, \citenamefont {Foss-Feig}, \citenamefont {Michalakis},
  \citenamefont {Gorshkov},\ and\ \citenamefont {Monroe}}]{Richerme2014}%
  \BibitemOpen
  \bibfield  {author} {\bibinfo {author} {\bibfnamefont {P.}~\bibnamefont
  {Richerme}}, \bibinfo {author} {\bibfnamefont {Z.-X.}\ \bibnamefont {Gong}},
  \bibinfo {author} {\bibfnamefont {A.}~\bibnamefont {Lee}}, \bibinfo {author}
  {\bibfnamefont {C.}~\bibnamefont {Senko}}, \bibinfo {author} {\bibfnamefont
  {J.}~\bibnamefont {Smith}}, \bibinfo {author} {\bibfnamefont
  {M.}~\bibnamefont {Foss-Feig}}, \bibinfo {author} {\bibfnamefont
  {S.}~\bibnamefont {Michalakis}}, \bibinfo {author} {\bibfnamefont {A.~V.}\
  \bibnamefont {Gorshkov}},\ and\ \bibinfo {author} {\bibfnamefont
  {C.}~\bibnamefont {Monroe}},\ }\bibfield  {title} {\bibinfo {title}
  {Non-local propagation of correlations in quantum systems with long-range
  interactions},\ }\href {\doibase 10.1038/nature13450} {\bibfield  {journal} {\bibinfo  {journal}
  {Nature}\ }\textbf {\bibinfo {volume} {511}},\ \bibinfo {pages} {198}
  (\bibinfo {year} {2014})}\BibitemShut {NoStop}%
\bibitem [{\citenamefont {Berry}\ and\ \citenamefont
  {Tabor}(1977)}]{Berry1977}%
  \BibitemOpen
  \bibfield  {author} {\bibinfo {author} {\bibfnamefont {M.~V.}\ \bibnamefont
  {Berry}}\ and\ \bibinfo {author} {\bibfnamefont {M.}~\bibnamefont {Tabor}},\
  }\bibfield  {title} {\bibinfo {title} {Level clustering in the regular
  spectrum},\ }\href {\doibase 10.1098/rspa.1977.0140} {\bibfield  {journal} {\bibinfo  {journal} {Proc.
  R. Soc. Lond. A. Math. Phys. Sc.}\ }\textbf {\bibinfo {volume} {356}},\
  \bibinfo {pages} {375} (\bibinfo {year} {1977})}\BibitemShut {NoStop}%
\bibitem [{\citenamefont {Bohigas}\ \emph {et~al.}(1984)\citenamefont
  {Bohigas}, \citenamefont {Giannoni},\ and\ \citenamefont
  {Schmit}}]{Bohigas1984}%
  \BibitemOpen
  \bibfield  {author} {\bibinfo {author} {\bibfnamefont {O.}~\bibnamefont
  {Bohigas}}, \bibinfo {author} {\bibfnamefont {M.-J.}\ \bibnamefont
  {Giannoni}},\ and\ \bibinfo {author} {\bibfnamefont {C.}~\bibnamefont
  {Schmit}},\ }\bibfield  {title} {\bibinfo {title} {Characterization of
  chaotic quantum spectra and universality of level fluctuation laws},\
  }\href {\doibase 10.1103/PhysRevLett.52.1} {\bibfield  {journal} {\bibinfo  {journal} {Phys. Rev. lett.}\
  }\textbf {\bibinfo {volume} {52}},\ \bibinfo {pages} {1} (\bibinfo {year}
  {1984})}\BibitemShut {NoStop}%
\bibitem [{\citenamefont {Prange}(1997)}]{Prange1997}%
  \BibitemOpen
  \bibfield  {author} {\bibinfo {author} {\bibfnamefont {R.~E.}\ \bibnamefont
  {Prange}},\ }\bibfield  {title} {\bibinfo {title} {The spectral form factor
  is not self-averaging},\ }\href {\doibase 10.1103/PhysRevLett.78.2280}
  {\bibfield  {journal} {\bibinfo  {journal} {Phys. Rev. Lett.}\ }\textbf
  {\bibinfo {volume} {78}},\ \bibinfo {pages} {2280} (\bibinfo {year}
  {1997})}\BibitemShut {NoStop}%
\bibitem [{\citenamefont {Schiulaz}\ \emph {et~al.}(2020)\citenamefont
  {Schiulaz}, \citenamefont {Torres-Herrera}, \citenamefont {P\'erez-Bernal},\
  and\ \citenamefont {Santos}}]{Schiulaz2020}%
  \BibitemOpen
  \bibfield  {author} {\bibinfo {author} {\bibfnamefont {M.}~\bibnamefont
  {Schiulaz}}, \bibinfo {author} {\bibfnamefont {E.~J.}\ \bibnamefont
  {Torres-Herrera}}, \bibinfo {author} {\bibfnamefont {F.}~\bibnamefont
  {P\'erez-Bernal}},\ and\ \bibinfo {author} {\bibfnamefont {L.~F.}\
  \bibnamefont {Santos}},\ }\bibfield  {title} {\bibinfo {title}
  {Self-averaging in many-body quantum systems out of equilibrium: Chaotic
  systems},\ }\href {\doibase 10.1103/PhysRevB.101.174312} {\bibfield
  {journal} {\bibinfo  {journal} {Phys. Rev. B}\ }\textbf {\bibinfo {volume}
  {101}},\ \bibinfo {pages} {174312} (\bibinfo {year} {2020})}\BibitemShut
  {NoStop}%
\bibitem [{\citenamefont {Torres-Herrera}\ \emph
  {et~al.}(2020{\natexlab{a}})\citenamefont {Torres-Herrera}, \citenamefont
  {De~Tomasi}, \citenamefont {Schiulaz}, \citenamefont {P{\'e}rez-Bernal},\
  and\ \citenamefont {Santos}}]{TorresHerrera2020a}%
  \BibitemOpen
  \bibfield  {author} {\bibinfo {author} {\bibfnamefont {E.~J.}\ \bibnamefont
  {Torres-Herrera}}, \bibinfo {author} {\bibfnamefont {G.}~\bibnamefont
  {De~Tomasi}}, \bibinfo {author} {\bibfnamefont {M.}~\bibnamefont {Schiulaz}},
  \bibinfo {author} {\bibfnamefont {F.}~\bibnamefont {P{\'e}rez-Bernal}},\ and\
  \bibinfo {author} {\bibfnamefont {L.~F.}\ \bibnamefont {Santos}},\ }\bibfield
   {title} {\bibinfo {title} {Self-averaging in many-body quantum systems out
  of equilibrium: Approach to the localized phase},\ }\href {\doibase 10.1103/PhysRevB.102.094310} {\bibfield
  {journal} {\bibinfo  {journal} {Phys. Rev. B}\ }\textbf {\bibinfo {volume}
  {102}},\ \bibinfo {pages} {094310} (\bibinfo {year}
  {2020}{\natexlab{a}})}\BibitemShut {NoStop}%
\bibitem [{\citenamefont {Torres-Herrera}\ \emph
  {et~al.}(2020{\natexlab{b}})\citenamefont {Torres-Herrera}, \citenamefont
  {Vallejo-Fabila}, \citenamefont {Mart{\'\i}nez-Mendoza},\ and\ \citenamefont
  {Santos}}]{TorresHerrera2020b}%
  \BibitemOpen
  \bibfield  {author} {\bibinfo {author} {\bibfnamefont {E.~J.}\ \bibnamefont
  {Torres-Herrera}}, \bibinfo {author} {\bibfnamefont {I.}~\bibnamefont
  {Vallejo-Fabila}}, \bibinfo {author} {\bibfnamefont {A.~J.}\ \bibnamefont
  {Mart{\'\i}nez-Mendoza}},\ and\ \bibinfo {author} {\bibfnamefont {L.~F.}\
  \bibnamefont {Santos}},\ }\bibfield  {title} {\bibinfo {title}
  {Self-averaging in many-body quantum systems out of equilibrium: Time
  dependence of distributions},\ }\href {\doibase 10.1103/PhysRevE.102.062126} {\bibfield  {journal} {\bibinfo
   {journal} {Phys. Rev. E}\ }\textbf {\bibinfo {volume} {102}},\ \bibinfo
  {pages} {062126} (\bibinfo {year} {2020}{\natexlab{b}})}\BibitemShut
  {NoStop}%
\bibitem [{\citenamefont {Torres-Herrera}\ and\ \citenamefont
  {Santos}(2015)}]{Torres2015}%
  \BibitemOpen
  \bibfield  {author} {\bibinfo {author} {\bibfnamefont {E.~J.}\ \bibnamefont
  {Torres-Herrera}}\ and\ \bibinfo {author} {\bibfnamefont {L.~F.}\
  \bibnamefont {Santos}},\ }\bibfield  {title} {\bibinfo {title} {Dynamics at
  the many-body localization transition},\ }\href
  {\doibase 10.1103/PhysRevB.92.014208} {\bibfield  {journal} {\bibinfo
  {journal} {Phys. Rev. B}\ }\textbf {\bibinfo {volume} {92}},\ \bibinfo
  {pages} {014208} (\bibinfo {year} {2015})}\BibitemShut {NoStop}%
\bibitem [{\citenamefont {Luitz}\ \emph {et~al.}(2016)\citenamefont {Luitz},
  \citenamefont {Laflorencie},\ and\ \citenamefont {Alet}}]{luitz2016}%
  \BibitemOpen
  \bibfield  {author} {\bibinfo {author} {\bibfnamefont {D.~J.}\ \bibnamefont
  {Luitz}}, \bibinfo {author} {\bibfnamefont {N.}~\bibnamefont {Laflorencie}},\
  and\ \bibinfo {author} {\bibfnamefont {F.}~\bibnamefont {Alet}},\ }\bibfield
  {title} {\bibinfo {title} {Extended slow dynamical regime close to the
  many-body localization transition},\ }\href {\doibase 10.1103/PhysRevB.93.060201} {\bibfield  {journal}
  {\bibinfo  {journal} {Phys. Rev. B}\ }\textbf {\bibinfo {volume} {93}},\
  \bibinfo {pages} {060201} (\bibinfo {year} {2016})}\BibitemShut {NoStop}%
\bibitem [{\citenamefont {Fischer}\ \emph {et~al.}(2016)\citenamefont
  {Fischer}, \citenamefont {Maksymenko},\ and\ \citenamefont
  {Altman}}]{Fischer2016}%
  \BibitemOpen
  \bibfield  {author} {\bibinfo {author} {\bibfnamefont {M.~H.}\ \bibnamefont
  {Fischer}}, \bibinfo {author} {\bibfnamefont {M.}~\bibnamefont
  {Maksymenko}},\ and\ \bibinfo {author} {\bibfnamefont {E.}~\bibnamefont
  {Altman}},\ }\bibfield  {title} {\bibinfo {title} {Dynamics of a
  many-body-localized system coupled to a bath},\ }\href {\doibase 10.1103/PhysRevLett.116.160401} {\bibfield
  {journal} {\bibinfo  {journal} {Phys. Rev. Lett.}\ }\textbf {\bibinfo
  {volume} {116}},\ \bibinfo {pages} {160401} (\bibinfo {year}
  {2016})}\BibitemShut {NoStop}%
\bibitem [{\citenamefont {Levi}\ \emph {et~al.}(2016)\citenamefont {Levi},
  \citenamefont {Heyl}, \citenamefont {Lesanovsky},\ and\ \citenamefont
  {Garrahan}}]{Levi2016}%
  \BibitemOpen
  \bibfield  {author} {\bibinfo {author} {\bibfnamefont {E.}~\bibnamefont
  {Levi}}, \bibinfo {author} {\bibfnamefont {M.}~\bibnamefont {Heyl}}, \bibinfo
  {author} {\bibfnamefont {I.}~\bibnamefont {Lesanovsky}},\ and\ \bibinfo
  {author} {\bibfnamefont {J.~P.}\ \bibnamefont {Garrahan}},\ }\bibfield
  {title} {\bibinfo {title} {Robustness of many-body localization in the
  presence of dissipation},\ }\href {\doibase 10.1103/PhysRevLett.116.237203} {\bibfield  {journal} {\bibinfo
  {journal} {Phys. Rev. Lett.}\ }\textbf {\bibinfo {volume} {116}},\ \bibinfo
  {pages} {237203} (\bibinfo {year} {2016})}\BibitemShut {NoStop}%
\bibitem [{\citenamefont {Torres-Herrera}\ and\ \citenamefont
  {Santos}(2017{\natexlab{a}})}]{TorresH2017}%
  \BibitemOpen
  \bibfield  {author} {\bibinfo {author} {\bibfnamefont {E.~J.}\ \bibnamefont
  {Torres-Herrera}}\ and\ \bibinfo {author} {\bibfnamefont {L.~F.}\
  \bibnamefont {Santos}},\ }\bibfield  {title} {\bibinfo {title} {Dynamical
  manifestations of quantum chaos: Correlation hole and bulge},\ }\href
  {\doibase 10.1098/rsta.2016.0434} {\bibfield  {journal} {\bibinfo
  {journal} {Philos. Trans. Royal Soc. A}\ }\textbf {\bibinfo {volume} {375}},\
  \bibinfo {pages} {20160434} (\bibinfo {year}
  {2017}{\natexlab{a}})}\BibitemShut {NoStop}%
\bibitem [{\citenamefont {Torres-Herrera}\ \emph {et~al.}(2018)\citenamefont
  {Torres-Herrera}, \citenamefont {Garc\'{\i}a-Garc\'{\i}a},\ and\
  \citenamefont {Santos}}]{Torres2018}%
  \BibitemOpen
  \bibfield  {author} {\bibinfo {author} {\bibfnamefont {E.~J.}\ \bibnamefont
  {Torres-Herrera}}, \bibinfo {author} {\bibfnamefont {A.~M.}\ \bibnamefont
  {Garc\'{\i}a-Garc\'{\i}a}},\ and\ \bibinfo {author} {\bibfnamefont {L.~F.}\
  \bibnamefont {Santos}},\ }\bibfield  {title} {\bibinfo {title} {Generic
  dynamical features of quenched interacting quantum systems: Survival
  probability, density imbalance, and out-of-time-ordered correlator},\ }\href
  {\doibase 10.1103/PhysRevB.97.060303} {\bibfield  {journal} {\bibinfo
  {journal} {Phys. Rev. B}\ }\textbf {\bibinfo {volume} {97}},\ \bibinfo
  {pages} {060303(R)} (\bibinfo {year} {2018})}\BibitemShut {NoStop}%
\bibitem [{\citenamefont {Torres-Herrera}\ and\ \citenamefont
  {Santos}(2019)}]{Torres2019}%
  \BibitemOpen
  \bibfield  {author} {\bibinfo {author} {\bibfnamefont {E.~J.}\ \bibnamefont
  {Torres-Herrera}}\ and\ \bibinfo {author} {\bibfnamefont {L.~F.}\
  \bibnamefont {Santos}},\ }\bibfield  {title} {\bibinfo {title} {Signatures of
  chaos and thermalization in the dynamics of many-body quantum systems},\
  }\href {\doibase 10.1140/epjst/e2019-800057-8} {\bibfield  {journal} {\bibinfo  {journal} {Eur. Phys. J.
  Spec. Top.}\ }\textbf {\bibinfo {volume} {227}},\ \bibinfo {pages} {1897}
  (\bibinfo {year} {2019})}\BibitemShut {NoStop}%
\bibitem [{\citenamefont {Lezama}\ \emph {et~al.}(2021)\citenamefont {Lezama},
  \citenamefont {Torres-Herrera}, \citenamefont {P{\'e}rez-Bernal},
  \citenamefont {Lev},\ and\ \citenamefont {Santos}}]{lezama2021}%
  \BibitemOpen
  \bibfield  {author} {\bibinfo {author} {\bibfnamefont {T.~L.}\ \bibnamefont
  {Lezama}}, \bibinfo {author} {\bibfnamefont {E.~J.}\ \bibnamefont
  {Torres-Herrera}}, \bibinfo {author} {\bibfnamefont {F.}~\bibnamefont
  {P{\'e}rez-Bernal}}, \bibinfo {author} {\bibfnamefont {Y.~B.}\ \bibnamefont
  {Lev}},\ and\ \bibinfo {author} {\bibfnamefont {L.~F.}\ \bibnamefont
  {Santos}},\ }\bibfield  {title} {\bibinfo {title} {Equilibration time in
  many-body quantum systems},\ }\href {\doibase 10.1103/PhysRevB.104.085117} {\bibfield  {journal} {\bibinfo
  {journal} {Phys. Rev. B}\ }\textbf {\bibinfo {volume} {104}},\ \bibinfo
  {pages} {085117} (\bibinfo {year} {2021})}\BibitemShut {NoStop}%
\bibitem [{\citenamefont {Siddiqui}(1962)}]{Siddiqui1962}%
  \BibitemOpen
  \bibfield  {author} {\bibinfo {author} {\bibfnamefont {M.~M.}\ \bibnamefont
  {Siddiqui}},\ }\bibfield  {title} {\bibinfo {title} {Some problems connected
  with {R}ayleigh distributions},\ }\href {https://api.semanticscholar.org/CorpusID:44230868} {\bibfield  {journal}
  {\bibinfo  {journal} {J. Res. Natl. Bur. Stand. (U. S.)}\ }\textbf {\bibinfo
  {volume} {66}},\ \bibinfo {pages} {167} (\bibinfo {year} {1962})}\BibitemShut
  {NoStop}%
\bibitem [{\citenamefont {Hilhorst}(2009)}]{Hilhorst2009}%
  \BibitemOpen
  \bibfield  {author} {\bibinfo {author} {\bibfnamefont {H.}~\bibnamefont
  {Hilhorst}},\ }\bibfield  {title} {\bibinfo {title} {Central limit theorem
  for correlated variables: Some critical remarks},\ }\href {\doibase 10.1590/S0103-97332009000400005} {\bibfield
  {journal} {\bibinfo  {journal} {Brazilian J. Phys.}\ }\textbf {\bibinfo
  {volume} {39}},\ \bibinfo {pages} {371} (\bibinfo {year} {2009})}\BibitemShut
  {NoStop}%
\bibitem [{\citenamefont {Aurich}\ and\ \citenamefont
  {Steiner}(1999)}]{Aurich1999}%
  \BibitemOpen
  \bibfield  {author} {\bibinfo {author} {\bibfnamefont {R.}~\bibnamefont
  {Aurich}}\ and\ \bibinfo {author} {\bibfnamefont {F.}~\bibnamefont
  {Steiner}},\ }\bibfield  {title} {\bibinfo {title} {Temporal quantum chaos},\
  }\href {\doibase 10.1142/S0217979299002459} {\bibfield  {journal} {\bibinfo  {journal} {Int. J. Mod. Phys.
  B}\ }\textbf {\bibinfo {volume} {13}},\ \bibinfo {pages} {2361} (\bibinfo
  {year} {1999})}\BibitemShut {NoStop}%
\bibitem [{\citenamefont {Kunz}(1999)}]{Kunz1999}%
  \BibitemOpen
  \bibfield  {author} {\bibinfo {author} {\bibfnamefont {H.}~\bibnamefont
  {Kunz}},\ }\bibfield  {title} {\bibinfo {title} {The probability distribution
  of the spectral form factor in random matrix theory},\ }\href {\doibase 10.1088/0305-4470/32/11/011}
  {\bibfield  {journal} {\bibinfo  {journal} {J. Phys. A Math. Gen.}\ }\textbf
  {\bibinfo {volume} {32}},\ \bibinfo {pages} {2171} (\bibinfo {year}
  {1999})}\BibitemShut {NoStop}%
\bibitem [{\citenamefont {Kunz}(2002)}]{Kunz2002}%
  \BibitemOpen
  \bibfield  {author} {\bibinfo {author} {\bibfnamefont {H.}~\bibnamefont
  {Kunz}},\ }\bibfield  {title} {\bibinfo {title} {Quantum dynamics and random
  matrix theory},\ }\href {\doibase 10.1142/S0217979202011731} {\bibfield  {journal} {\bibinfo  {journal}
  {Int. J. Mod. Phys. B}\ }\textbf {\bibinfo {volume} {16}},\ \bibinfo {pages}
  {2003} (\bibinfo {year} {2002})}\BibitemShut {NoStop}%
\bibitem [{\citenamefont {Serbyn}\ and\ \citenamefont
  {Moore}(2016)}]{Serbyn2016}%
  \BibitemOpen
  \bibfield  {author} {\bibinfo {author} {\bibfnamefont {M.}~\bibnamefont
  {Serbyn}}\ and\ \bibinfo {author} {\bibfnamefont {J.~E.}\ \bibnamefont
  {Moore}},\ }\bibfield  {title} {\bibinfo {title} {Spectral statistics across
  the many-body localization transition},\ }\href {\doibase 10.1103/PhysRevB.93.041424} {\bibfield  {journal}
  {\bibinfo  {journal} {Phys. Rev. B}\ }\textbf {\bibinfo {volume} {93}},\
  \bibinfo {pages} {041424} (\bibinfo {year} {2016})}\BibitemShut {NoStop}%
\bibitem [{\citenamefont {Torres-Herrera}\ and\ \citenamefont
  {Santos}(2017{\natexlab{b}})}]{Torres2017}%
  \BibitemOpen
  \bibfield  {author} {\bibinfo {author} {\bibfnamefont {E.~J.}\ \bibnamefont
  {Torres-Herrera}}\ and\ \bibinfo {author} {\bibfnamefont {L.~F.}\
  \bibnamefont {Santos}},\ }\bibfield  {title} {\bibinfo {title} {Extended
  nonergodic states in disordered many-body quantum systems},\ }\href
  {\doibase 10.1002/andp.201600284} {\bibfield  {journal} {\bibinfo
  {journal} {Ann. Phys. (Berlin)}\ }\textbf {\bibinfo {volume} {529}},\
  \bibinfo {pages} {1600284} (\bibinfo {year}
  {2017}{\natexlab{b}})}\BibitemShut {NoStop}%
\bibitem [{\citenamefont {Tikhonov}\ and\ \citenamefont
  {Mirlin}(2021)}]{Tikhonov2021}%
  \BibitemOpen
  \bibfield  {author} {\bibinfo {author} {\bibfnamefont {K.~S.}\ \bibnamefont
  {Tikhonov}}\ and\ \bibinfo {author} {\bibfnamefont {A.~D.}\ \bibnamefont
  {Mirlin}},\ }\bibfield  {title} {\bibinfo {title} {Eigenstate correlations
  around the many-body localization transition},\ }\href {\doibase 10.1103/PhysRevB.103.064204} {\bibfield
  {journal} {\bibinfo  {journal} {Phys. Rev. B}\ }\textbf {\bibinfo {volume}
  {103}},\ \bibinfo {pages} {064204} (\bibinfo {year} {2021})}\BibitemShut
  {NoStop}%
\bibitem [{\citenamefont {L{\"u}schen}\ \emph {et~al.}(2017)\citenamefont
  {L{\"u}schen}, \citenamefont {Bordia}, \citenamefont {Scherg}, \citenamefont
  {Alet}, \citenamefont {Altman}, \citenamefont {Schneider},\ and\
  \citenamefont {Bloch}}]{Luschen2017}%
  \BibitemOpen
  \bibfield  {author} {\bibinfo {author} {\bibfnamefont {H.~P.}\ \bibnamefont
  {L{\"u}schen}}, \bibinfo {author} {\bibfnamefont {P.}~\bibnamefont {Bordia}},
  \bibinfo {author} {\bibfnamefont {S.}~\bibnamefont {Scherg}}, \bibinfo
  {author} {\bibfnamefont {F.}~\bibnamefont {Alet}}, \bibinfo {author}
  {\bibfnamefont {E.}~\bibnamefont {Altman}}, \bibinfo {author} {\bibfnamefont
  {U.}~\bibnamefont {Schneider}},\ and\ \bibinfo {author} {\bibfnamefont
  {I.}~\bibnamefont {Bloch}},\ }\bibfield  {title} {\bibinfo {title}
  {Observation of slow dynamics near the many-body localization transition in
  one-dimensional quasiperiodic systems},\ }\href {\doibase 10.1103/PhysRevLett.119.260401} {\bibfield  {journal}
  {\bibinfo  {journal} {Phys. Rev. Lett.}\ }\textbf {\bibinfo {volume} {119}},\
  \bibinfo {pages} {260401} (\bibinfo {year} {2017})}\BibitemShut {NoStop}%
\bibitem [{\citenamefont {Willemain}\ and\ \citenamefont
  {Desautels}(1993)}]{Thomas1993}%
  \BibitemOpen
  \bibfield  {author} {\bibinfo {author} {\bibfnamefont {R.}~\bibnamefont
  {Willemain}}\ and\ \bibinfo {author} {\bibfnamefont {A.}~\bibnamefont
  {Desautels}},\ }\bibfield  {title} {\bibinfo {title} {A method to generate
  autocorrelated uniform random numbers},\ }\href
  {\doibase 10.1080/00949659308811469} {\bibfield  {journal} {\bibinfo
  {journal} {J. Stat. Comput. Simul.}\ }\textbf {\bibinfo {volume} {45}},\
  \bibinfo {pages} {23} (\bibinfo {year} {1993})}\BibitemShut {NoStop}%
\bibitem [{\citenamefont {Vallejo-Fabila}\ and\ \citenamefont
  {Torres-Herrera}(2022)}]{Vallejo2022}%
  \BibitemOpen
  \bibfield  {author} {\bibinfo {author} {\bibfnamefont {I.}~\bibnamefont
  {Vallejo-Fabila}}\ and\ \bibinfo {author} {\bibfnamefont {E.~J.}\
  \bibnamefont {Torres-Herrera}},\ }\bibfield  {title} {\bibinfo {title}
  {Effects of autocorrelated disorder on the dynamics in the vicinity of the
  many-body localization transition},\ }\href {\doibase 10.1103/PhysRevB.106.L220201} {\bibfield  {journal}
  {\bibinfo  {journal} {Phys. Rev. B}\ }\textbf {\bibinfo {volume} {106}},\
  \bibinfo {pages} {L220201} (\bibinfo {year} {2022})}\BibitemShut {NoStop}%
\bibitem [{\citenamefont {Chirikov}\ and\ \citenamefont
  {Shepelyansky}(1995)}]{Chirikov1995}%
  \BibitemOpen
  \bibfield  {author} {\bibinfo {author} {\bibfnamefont {B.}~\bibnamefont
  {Chirikov}}\ and\ \bibinfo {author} {\bibfnamefont {D.}~\bibnamefont
  {Shepelyansky}},\ }\bibfield  {title} {\bibinfo {title} {Shnirelman peak in
  level spacing statistics},\ }\href {\doibase 10.1103/PhysRevLett.74.518} {\bibfield  {journal} {\bibinfo
  {journal} {Phys. Rev. Lett.}\ }\textbf {\bibinfo {volume} {74}},\ \bibinfo
  {pages} {518} (\bibinfo {year} {1995})}\BibitemShut {NoStop}%
\bibitem [{\citenamefont {Wegner}(1980)}]{Wegner1980}%
  \BibitemOpen
  \bibfield  {author} {\bibinfo {author} {\bibfnamefont {F.}~\bibnamefont
  {Wegner}},\ }\bibfield  {title} {\bibinfo {title} {Inverse participation
  ratio in $2+\varepsilon$ dimensions},\ }\href {\doibase 10.1007/BF01325284} {\bibfield  {journal}
  {\bibinfo  {journal} {Z. Phys. B Con. Mat.}\ }\textbf {\bibinfo {volume}
  {36}},\ \bibinfo {pages} {209} (\bibinfo {year} {1980})}\BibitemShut
  {NoStop}%
\end{thebibliography}
\end{document}